\documentclass{aa}  

\usepackage{graphicx}
\usepackage{placeins}
\usepackage{txfonts}
\usepackage{ulem}
\usepackage{pdflscape}
\usepackage{amsmath}
\usepackage[caption=false]{subfig}
\usepackage{natbib}

\usepackage{mathabx}
\usepackage[colorlinks=true,citecolor=blue]{hyperref}

\hypersetup{colorlinks,allcolors=blue}

\newcommand{\MSun}{M$_{\odot}$\xspace}

\newcommand{\kms}{km\, s$^{-1}$\xspace}

\begin{document}

   \title{The Corona-Australis star-forming region:}

   \subtitle{New insights on its formation history from detailed stellar and disk analysis}

   \author{E. Rigliaco\inst{1}, R. Gratton \inst{1} and V. Nascimbeni \inst{1}}
   \institute{\inst{1}INAF - Osservatorio Astronomico di Padova, ITALY\\
              \email{elisabetta.rigliaco@inaf.it}
             }

   \date{Received 14 April 2025; accepted 25 July 2025}
 
  \abstract
   { The star-forming complex of Corona Australis (CrA) is one of the closest and most isolated molecular clouds. It belongs to a chain of clusters that show age gradients with distance from the galactic plane.  }
   { We aim to provide suggestions regarding its formation history by examining the stellar and disk populations, the stellar multiplicity, and the interstellar absorption.}
   { We made a census of stars and disks using Gaia DR3 and infrared data. Interstellar absorption in the direction of each star was derived by comparing spectral types from the literature and Gaia colors. Stellar multiplicity analysis accounts for both direct observation of visual companions (Gaia data and high-contrast imaging) and indirect detection of the presence of companions (eclipsing binaries, spectroscopic binaries, and astrometry). The properties of the disks were obtained from the slopes of the spectral energy distributions. }
    {As found in previous studies, the CrA complex can be divided into two regions: a younger region (CrA-Main: 3$\pm$1~Myr) and an older one (CrA-North: 6.7$\pm$0.3~Myr), which are slightly younger than previously thought. Moreover, while CrA-Main still appears bound to the gas, CrA-North is unbound and expanding. The stars that belong to CrA-North were in the most compact configuration 3.72~Myr ago. At that time, CrA-Main and CrA-North were also much closer to each other than they appear now. The fraction of disk-bearing stars is higher in CrA-Main than in CrA-North, as also expected due to the younger age of CrA-Main.}
    {We propose a formation history scenario for the CrA-complex. It started between 15 and 18~Myr ago with supernovae (SNe) explosions in the Upper Centaurus-Lupus complex, followed by a quiescent phase with little star formation. A star formation episode $\sim$7~Myr ago formed CrA-North stars. About 3.7~Myr ago, a second SN explosion south of CrA-North triggered star formation in CrA-Main. This last SN might have been the origin of the pulsar RX J1856.5-3754.}


   \maketitle

\section{Introduction}
\label{section:introduction}

Star formation is a continuous process that lasts for periods as long as 10$^7$ years in many molecular cloud complexes. However, tracing the star formation history of these molecular clouds is not trivial and it is not possible without precise parallaxes, proper motions, and radial velocity measurements such as those provided by the European Space Agency (ESA) Gaia satellite. The scenario that has emerged in recent years is that most young clusters in the solar neighborhood originated in three compact massive star-forming complexes \citep{Swiggum2024}. Moreover, it is now clear that large OB associations (e.g., the Scorpius-Centaurus - Sco-Cen - OB association analyzed by \citealt{Ratzenbock2023b, Ratzenbock2023a}) contain chains as long as hundreds of parsec  of continuous clusters that show well-defined age gradients, from massive older clusters to smaller younger clusters. Reconstructing the history of these star-forming regions is an essential step in gaining insight into the star formation process itself. 

The Corona Australis (CrA) star formation complex represents one of the ideal locations to conduct a detailed analysis of the stellar and disk properties of the clusters related to the Sco-Cen OB association. Indeed, contrary to other subgroups of the Sco-Cen, CrA is very well isolated, which enables a very accurate selection of the stars belonging to it. Located $\sim$18$^{\degree}$ south of the galactic plane, it is at a distance of about 150 pc \citep{Zucker2020}. Its densest region harbors a deeply embedded cluster, the Coronet cluster \citep{Taylor1984,Neuh2008}, with ongoing and/or recent star formation, and is centered around R CrA, which was the first variable star noticed in the CrA constellation \citep{Reynolds1916}. 

The CrA region has recently been analyzed by several authors. 
 Using Gaia Data Release 2 (DR2, \citealt{Gaia2016}) data, \citet{Galli2020} identified two kinematically and spatially distinct subgroups of young stars: the "off-cloud"  population, which is located toward the galactic north of the complex, and the "on-cloud" population, which is concentrated close to the densest regions of CrA, namely the Coronet cluster. 
Furthermore, \citet {Esplin2022} compiled a catalog of young stellar objects (YSOs) in CrA that contain off- and on-cloud stars. However, they made a different group classification with respect to the off-cloud and on-cloud definition proposed by \citet{Galli2020} and call the groups Upper Corona Australis (which contains the majority of the stars and roughly corresponds to the off-cloud population) and Coronet Cluster (which roughly corresponds to the on-cloud population).
Moreover, CrA is one of the clusters identified by \citet{Ratzenbock2023a}, who reconstructed the star formation history of the Sco-Cen OB association using a clustering algorithm called significance mode analysis (SigMA). The algorithm interprets the density peaks separated by dips as significant clusters. 
The application of SigMA to Gaia Data Release 3 (DR3) data \citep{Gaia2016, Gaia2023a} on stars in and around the Sco-Cen association led \citet{Ratzenbock2023a} to the discovery of multiple clusters. Among these clusters, they identified CrA-Main and CrA-North clusters, which coincide very well with the on-cloud populations, which are concentrated around the densest core, and off-cloud populations, which are located in the north of the dark clouds,  identified by \citet{Galli2020}. 

Follow-up studies on the expanding 3D velocity and position conducted by \citet{Posch2023} show that CrA-Main, which lies farther south of the galactic plane, moves away from the galactic plane faster than CrA-North. This is in contrast with what one would expect if there were no external forces in the galactic potential (which would slow down after crossing the galactic plane). Moreover, the shape of the CrA molecular cloud suggests the influence of a force originating from the direction of the Sco-Cen OB association. Putting together these hints, \citet{Posch2023} suggest that the observed gradient is caused by stellar feedback, namely the explosion of two supernovae (SNe) that have pushed and accelerated the gas away from the galactic plane. 

In this study, we develop these recent findings. The aim of this paper is to reconstruct the star formation history of this nearby and isolated star-forming complex by means of examining the stellar and disk population, the stellar multiplicity, and the interstellar absorption.
We redefine the stellar sample by including stars that are bright in the infrared and that were excluded from the comprehensive analysis of \citet{Ratzenbock2023a}. In Section~\ref{section:stellar_properties} we define the stellar sample by performing an accurate stellar census and we discuss our methods for considering the interstellar absorption and the stellar multiplicity. In Section~\ref{section:disk} we present the disk properties of the stars belonging to CrA-Main and CrA-North. In Section~\ref{section:discussion} we discuss our findings and summarize them in Section~\ref{section:conclusions}.

\section{Datasets and stellar census} 
\label{section:data}
In this paper, we consider a total of 442 stars in the CrA-complex, 134 belonging to CrA-Main and 308 belonging to CrA-North (all listed in Table~\ref{tab:main_TAB_1}, with increasing identification number Run-ID from 1 to 442).
The starting point of the CrA complex stellar census is the stellar clustering selection performed by \citet{Ratzenbock2023a} from {\it Gaia} DR3 data. They considered in their study three SigMA clusters in the CrA region. A first cluster (SigMA number 29 in their list) is projected on top of the CrA molecular cloud and coincides with the embedded Coronet clusters. They named it the CrA-Main group. At the galactic north of this group, they identified a second and more extended group (SigMA number 30 in their list), called CrA-North, which was already discussed in \citet{Galli2020} and \citet{Esplin2022}. Additionally, they identified a third group to the galactic northwest of the two other clusters, apparently building a bridge to the main body of Sco-Cen. This group was named Scorpio-Sting and is not considered in this work since its projected location is further out with respect to the other two clusters. Following \citet{Ratzenbock2023a}, we call CrA-Main and CrA-North the two groups of stars in the CrA complex throughout this paper. We call CrA-Cloud the interstellar cloud associated with the CrA-complex and CrA-cloud core the denser part of this cloud associated with CrA-Main.

According to \citet{Ratzenbock2023a}, CrA-Main and CrA-North count together 447 members (96 belonging to CrA-Main and 351 belonging to CrA-North). We started from their lists, but we revised their counts and membership as follows. First, four stars are accounted for in both CrA-Main and CrA-North, and they likely belong to CrA-Main. 
Second, there are 36 pairs and one triad of objects that have a projected separation below 60 arcsec, which corresponds to about 9000 au at the distance of the CrA-complex, we classify these objects as multiple systems (see also Sect.~\ref{subsection:binary}). Of these systems, 7 belong to CrA-Main, 24 to CrA-North (including the triple system), while for six systems, the two components were assigned to different groups by \citet{Ratzenbock2023a}. In this study, we assume that these six systems belong to CrA-Main because it has a more restrictive membership definition, and deviation of astrometry from median motion of the group is expected to be statistically higher for components of multiple systems because of the orbital motion around the center of mass of the system. 

We also considered 46 additional sources that are bright in the infrared \citep{Peterson2011} but were not considered by \citet{Ratzenbock2023a} either because they are faint or do not have entries in Gaia DR3, or have large astrometric errors (e.g., R CrA and T CrA) because they are deeply embedded in the CrA-cloud core. However, six of these sources are only detected at wavelengths longer than 3.6~$\mu$m and might be starless cores rather than stars. We did not consider them any further. The sources missing Gaia data lack high-precision astrometry measurements; we attributed 35 of them to CrA-Main and 5 to CrA-North on the basis of their projected position. We also re-examined the position of all stars in the color-magnitude diagram (CMD), and concluded that 16 stars (11 originally attributed to CrA-North and five to CrA-Main) are clearly much older than the others and are not actually members of the CrA complex. In addition, after examining their position in the position versus proper motion space, we attributed to CrA-Main eight sources that were considered members of CrA-North by \citet{Ratzenbock2023a}, and vice versa we attributed to CrA-North four stars that were originally attributed to CrA-Main. 

Hence, we considered 122 systems (134 stars) for the CrA-Main group and 308 stars for the CrA-North group. They are all reported in Table~\ref{tab:main_TAB_1} and \ref{tab:main_TAB_2}. 
For these objects, we collected data from the following sources: \\
Gaia data. We use Gaia DR3 data \citep{Gaia2023a} to retrieve astrometric data (proper motions (PM) in right ascension and declination, distances), photometric data (G-band magnitudes, Bp-Rp colors), and the RUWE (Renormalized Unit Weight Error) value. 109 systems belonging to Cra-Main have at least one entry in the Gaia DR3 catalog; the 308 systems for the CrA-North group are all listed in Gaia DR3.\\
2MASS data. Near-infrared photometric data (J--, H--, and K--band magnitudes) available for 121 stars in CrA-Main and 307 stars in CrA-North, have been obtained from the Two Micron All Sky Survey (2MASS) archive \citep{Skrutskie2006}. \\
WISE data.  AllWISE Source Catalog of the Wide-field Infrared Survey Explorer (WISE; \citealt{Wright2010}) was used to retrieve far-infrared photometric data that have been collected for 121 stars in CrA-Main and 298 stars in CrA-North from \citet{Peterson2011} and \citet{Sicilia-Aguilar2013}. \\
ALMA data. ALMA 1.3 mm continuum observations from \citet{Cazzoletti2019} have been used to retrieve the dust mass of 21 disks in the CrA-Main group, as listed in Table~\ref{tab:main_TAB_2}. 

Finally, we possibly considered as a member of CrA-North the very well known pulsar RX J1856.5-3754 (one of the magnificent seven:  \citealt{Kaplan2008}), which coincides in position and distance ($167^{+18}_{-15}$ pc \citealt{Kaplan2007} with the CrA-complex and has an estimated spin-down age of 3.72$\pm$0.06~Myr \citep{DeGrandis2022, Bogdanov2024}\footnote{We also notice that the position angle of $PA=97\pm 1$ degree of the bow-shock or cometary nebula observed by \citet{vanKerkwijk2001} is similar to the position angle of the relative proper motion of RX J1856.5-3754 with respect to CrA-Main ($95.7\pm 0.1$ degree). For comparison, the proper motion direction of RX J1856.5-3754 in the sky is $PA=100.25\pm 0.1$ degrees. This agrees fairly well with the assumption that RX J1856.5-3754 is moving within the CrA-cloud.}. However, the high proper motion measured for this object by \citet{Walter2002} implies that adopting this age it originated very far from its current position, and it is then possibly unrelated to the CrA-complex. We discuss this point again in Section~\ref{section:conclusions}.

\section{Results}
\label{section:stellar_properties}

\subsection{Interstellar absorption in the direction of CrA complex}
\label{subsection:absorption}

A major difficulty in the analysis of the CrA complex is related to the strong variability of the interstellar absorption toward individual objects in CrA-Main. There are various maps of interstellar absorption toward the CrA complex \citep{Cambresy1999, Dobashi2005}, but they do not have the resolution required to attribute an appropriate absorption to individual stars in the crowded region of CrA-Main. Higher-resolution maps of dust thermal emission can be obtained using results from the Herschel \citep{Bresnahan2018} and Planck satellites. However, they need to be calibrated in terms of extinction, and in addition they refer to the integrated emission in any given direction and are then appropriate only for objects that are behind the CrA-cloud, while the targets considered in this paper might be in front of or embedded within it. 
The values of the interstellar absorption A$_G$ appropriate to each star obtained by comparing their spectral types (as listed in \citealt{Esplin2022}), with the observed Gaia $BP-RP$ colors, are reported in Table~\ref{tab:main_TAB_2}. 
Spectral types are available for 110 stars in CrA-Main and 241 stars in CrA-North. However, 8 of the stars in CrA-Main do not have reliable Gaia colors, either because they are not in Gaia or because they have $G>20$. We used NIR photometry for them (namely, 2MASS J-H color). For the remaining stars (14 stars in CrA-Main and 67 stars in CrA-North) we adopted a calibration in terms of absorption of the galactic dust emission given by the Planck satellite; this calibration was obtained using results for stars with spectrum classification. All these stars but three are in regions with low interstellar absorption.
For the reddening law, we adopted the \citet{Cardelli1989} absorption law with a value of $R_V=4$ for the ratio between total and selective absorption. We checked that values in the range 3--5 for this parameter influence only marginally the results of this paper (e.g., the definition of disk-bearing stars or the ages obtained by comparison of the CMD with isochrones). 

\subsection{Stellar multiplicity}
\label{subsection:binary}

We search for companions to the stars in our sample following the approach adopted in previous papers by \citet{Gratton2023a, Gratton2023b, Gratton2024, Gratton2025}. The search is based on observation of visual companions (mainly from Gaia and high-contrast imaging) and indirect detection of the presence of companions from photometry (eclipsing binaries), RV (spectroscopic binaries) and astrometry (mainly from Gaia). The CrA complex is farther and younger than most of the samples considered in the papers cited above, and most of its components are faint. This results in a lower level of completeness. More specifically, the search is virtually insensitive to planetary companions and only moderately sensitive to stellar companions at a separation shorter than about a hundred au.

\subsubsection{Visual binaries}\label{vis_bin}

\begin{figure}
   \centering
   \includegraphics[width=8.0cm]{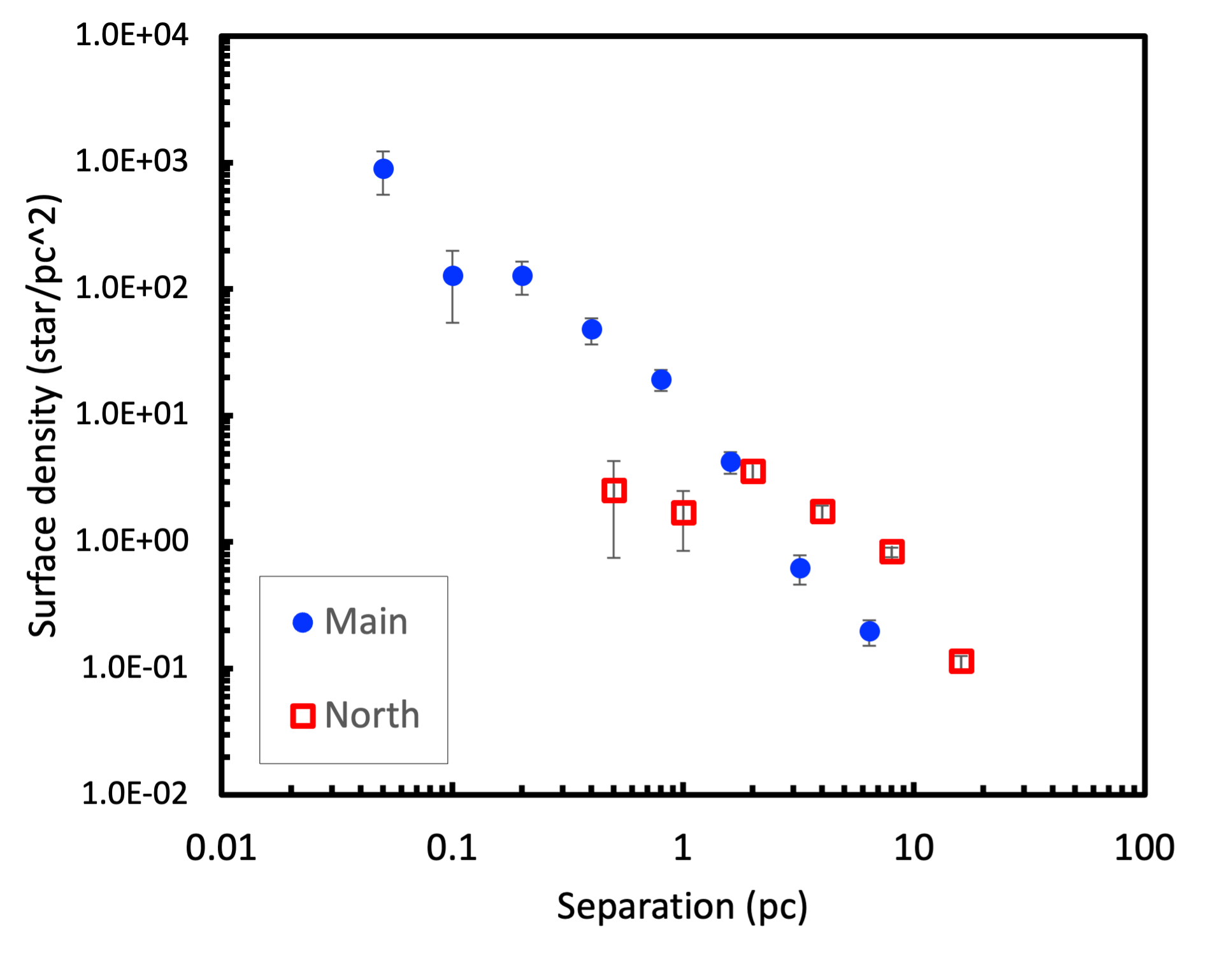}
      \caption{Surface density of stars in the CrA-complex as a function of distance from the center for CrA-Main (filled blue circles) and CrA-North (open red squares).  }
         \label{Fig:surface_density}
\end{figure}

Wide visual companions (separation $>0.7$ arcsec) may be detected either as separate entries in Gaia DR3 (but for stars with very high absorption), or by high-contrast imaging that is available for 26\% of the stars in CrA-Main and 5\% of the stars in CrA-North. 
Given the young age of the CrA complex, Gaia DR3 can reveal companions down to about 0.015~\MSun, corresponding to stars and brown dwarfs with separation larger than a few hundred~au. 
High-contrast imaging can also be sensitive to most massive planetary companions, but only in the separation range from a few tens to about 1000~au. 

We considered as potential physical companions all point sources that are members of the CrA complex and are within 60 arcsec (about 9000~au) from a putative primary. However, some of these potential companions might be members of the CrA complex that are not physically associated with their putative primaries but simply projected close to them. To estimate the probability that this occurs, we computed the surface density of stars in the CrA complex as a function of the projected distance from the center. We considered the CrA-Main and CrA-North groups separately (see Figure \ref{Fig:surface_density}). For CrA-Main, we adopted as center of the group the position of R~CrA (the most massive member) and for CrA-North the average position of all stars. We adopted two different definitions for the center of the Main and North group because of the potential incompleteness of member detection in CrA-Main; in fact, due to the presence of the very extincted region around the Coronet cluster we cannot observe all the stars and use the average position of all stars as done for CrA-North. 
In the case of CrA-Main the slope of the different points is well described by a power law with an exponent equal to $-$1.66, which is similar to the typical exponent of the power law for the analogous quantity in the interstellar medium for star-forming regions ($-$1.5: \citealt{Beuther2024}\footnote{This typical slope is obtained for cores of 12 arcsec, that at a typical distance of 3 kpc is about 0.2 pc.}). 
The surface density appears almost constant at about 1.2 star/pc$^2$ for CrA-North up to about 4 pc (about 1.5 degrees from the center) and declines at larger distances.

We used this surface density to estimate the probability that each of the detected objects within 60~arcsec is a physical companion. The probability of a casual alignment to another member of the CrA complex within 60 arcsec is significant in the central region of CrA-Main, while it is low ($\leq 0.7$\%) for CrA-North. 
In particular, we found that the brown dwarf 2MASS J19015374-3700339 (Run-ID 87) has a probability higher than 50\% to be casually aligned to 2MASS J19014936-3700285 (Run-ID 80 - that lies at a projected separation of 53~arcsec).  
We did not consider this object as a physical companion. 
The probability is 15\% for 2MASS J19012872-3659317 (Run-ID 61) which is at 38 arcsec from 2MASS J19012717-3659085 (Run-ID 60); 11\% for Gaia DR3 6731197442076732928 (Run-ID 67) at 30 arcsec from V709 CrA (Run-ID 66, itself a binary). 
It is possible that these last objects are not bound to their primaries, but we consider them as companions in our analysis. The probability of a chance alignment with another star in the CrA complex is $<5$\% for all remaining visual companions.
We also noticed that the astrometric solution may be missing for faint sources very close to brighter objects in the Gaia DR3 catalog. We then additionally considered as candidate companions objects listed in Gaia projected within 2 arcsec of each star but lacking an astrometric solution. We found 19 such objects. Though far from the galactic plane, the CrA-complex is projected toward a rich star background field, with contributions from the outer regions of the bulge (the line of sight passes about 2 kpc south of the galactic center), the outskirts of the Sagittarius dwarf galaxy (whose center is at RA=283.83 and Dec=-30.55), and the globular cluster NGC 6723 (center at RA=284.89 and Dec=-36.63). For each of the candidate companions, we then computed the probability that it is a background source by chance projected close to a star in the CrA-complex from the surface density of sources listed in Gaia DR3 brighter than the candidate companion in a region centered on the star; the radius of this region was adapted to have about 1500 Gaia sources brighter than the candidate companion. Using these surface densities, we found that the probability of a casual alignment is higher than 0.002 for 14 of these objects. These candidate companions were discarded because there is a $>50$\% probability of finding at least one chance alignment in the sample analyzed in this paper of the CrA complex selected from Gaia DR3, even if the probability for an individual object is as low as that. The remaining five high-probability companions were kept in our analysis.

\subsubsection{Eclipsing binaries}

Some eclipsing binaries (EB) are known in the CrA-complex. \citet{Vanko2013} analyzed TY CrA. In addition, detailed studies showed that both R CrA \citep{Sissa2019} and T CrA \citep{Rigliaco2023} are multiple systems where occultation by the disk causes periodic dimming of the components during their orbital motion around the center of mass of the system. All these objects are in CrA-Main.

We searched for EBs within the Transiting Exoplanet Survey Satellite (TESS: \citealt{Ricker2015}) dataset. We found short-cadence observations for 43 targets and additional 324 targets in the long-cadence mode. The light curves from these last observations were extracted using various pipelines: CDIPS \citep{Bouma2019}, PATHOS \citep{Nardiello2019}, QLP \citep{Huang2020}, SPOC \citep{Caldwell2020}.  
We found two EB candidates among the objects observed in long-cadence modes, both of them in CrA-North. 
TIC~253714131 (Run-ID 378) is a detached binary with an orbital period of 2.273~d. Photometry, spectral type, and light curve are all compatible with a very reddened ($A_G=3.2$ mag\footnote{Given its position quite far from the CrA-cloud core, this high reddening is likely due to the circumstellar disk, whose presence is indicated by the IR excess, see Section \ref{section:disk}.}) 8~Myr old equal-mass system in circular orbit with two components each having a mass of 0.177 \MSun, seen at an inclination of $i=88.6$ degrees. 
TIC~405363950 (Run-ID 213) is a detached binary with an orbital period of 1.18797458 d and a high eccentricity, which is a very unusual combination \citep{Shivvers2014}. Photometry and spectral type indicate masses of 0.077 and 0.060 \MSun for the two components, again assuming an age of 8 Myr. 

\subsubsection{Spectroscopic binaries}

We inspected the $S_B{^9}$ (\citealt{Pourbaix2004}, \citealt{Tokovinin2018})  and Gaia DR3 binary \citep{Gaia2023b} catalogs for spectroscopic and spectrophotometric binaries. Very few spectroscopic binaries are known in the CrA complex due to the scarce data available (due to the faintness of the targets) and large errors in RVs for very young stars. 
RV series are available for only 10\% of the stars in CrA-Main and 14\% of those in CrA-North. The parameter values for the tertiary spectroscopic of TY CrA are from \citet{Tokovinin2018}, and those for 2MASS J18430029-3416068 (Run-ID 304) are from \citet{Gaia2023b}.

\subsubsection{Astrometric binaries}

We inspected various catalogs looking for astrometric binaries based on Gaia data: nss\_two\_body\_orbit \citep{Gaia2023b}, nss\_acceleration\_astro (\citealt{Gaia2023b}, \citealt{Holl2023}) 
and found no entries for the stars in the CrA complex.  
We considered Proper Motion acceleration (PMa) from \citet{Kervella2022}. The PMa is the difference between the proper motion in Gaia DR3 (baseline of 34 months) and that determined using the position at Hipparcos (1991.25) and Gaia DR3 (2016.0) epochs. 
This quantity is available for only seven stars in the CrA-complex. 
PMa is sensitive to binaries with a projected separation between 1 and 100 au. We considered any value of PMa with a signal-to-noise ratio $S/N>3$ as an indication of the presence of companions and found three cases: HR 7169, HR 7170 (whose PMa is due to the visual secondary considered by \citealt{Kohler2008, Oudmaijer2010}), and HIP 93689 (HD 177076).

We also considered the re-normalized unit weight error (RUWE) as an indication of binarity. This parameter is an indication of the goodness of the 5-parameter solution found by Gaia \citep{Lindegren2018}. \citet{Belokurov2020} showed that RUWE values $>$ 1.4 suggest the presence of a companion, at least for stars that are not too bright ($G>4$) and saturated in the Gaia scans. This method is sensitive to systems that have periods from a few months to a decade \citep{Penoyre2022}. The RUWE parameter is available for all stars in CrA-North and for 81\% of the stars in CrA-Main.

\subsubsection{Parameters for the components}
\label{subsubsection:parameters}

When possible, the semimajor axis of the orbit and the mass of the companions for binary systems were taken from sources in the literature. When this was not possible, we followed the methods considered in \citet{Gratton2023a, Gratton2024, Gratton2025}, briefly summarized in the following.

We derived the masses of resolved visual binaries from the photometry of the individual stars and the isochrones by \citet{Baraffe2015}, assuming ages of 3 Myr for CrA-Main and 8 Myr\footnote{This is the closest match in the grid by \citet{Baraffe2015} to the age of 6.9 Myr for stars in CrA-North we obtained from the location of stars in the CMD; see Section \ref{subsection:ages}.} for CrA-North. 
For unresolved systems, the sum of the masses was made compatible with the apparent $G$-magnitude of the system, using the mass-luminosity relation for the Gaia $G$-band appropriate for the age of CrA-Main and CrA-North. We assumed that the semimajor axis is equal to the projected separation divided by the parallax. On average, this corresponds to the thermal eccentricity distribution considered by \citet{Ambartsumian1937} of $f(e)=2 e$ (see \citealt{Brandeker2006}). Uncertainties in the masses derived using these recipes are small (well below 10\%), while those for the semimajor axes are about 40\% (see Figure A.1 in \citealt{Brandeker2006}).

\begin{figure}[ht]
   \centering
   \includegraphics[width=8.4cm]{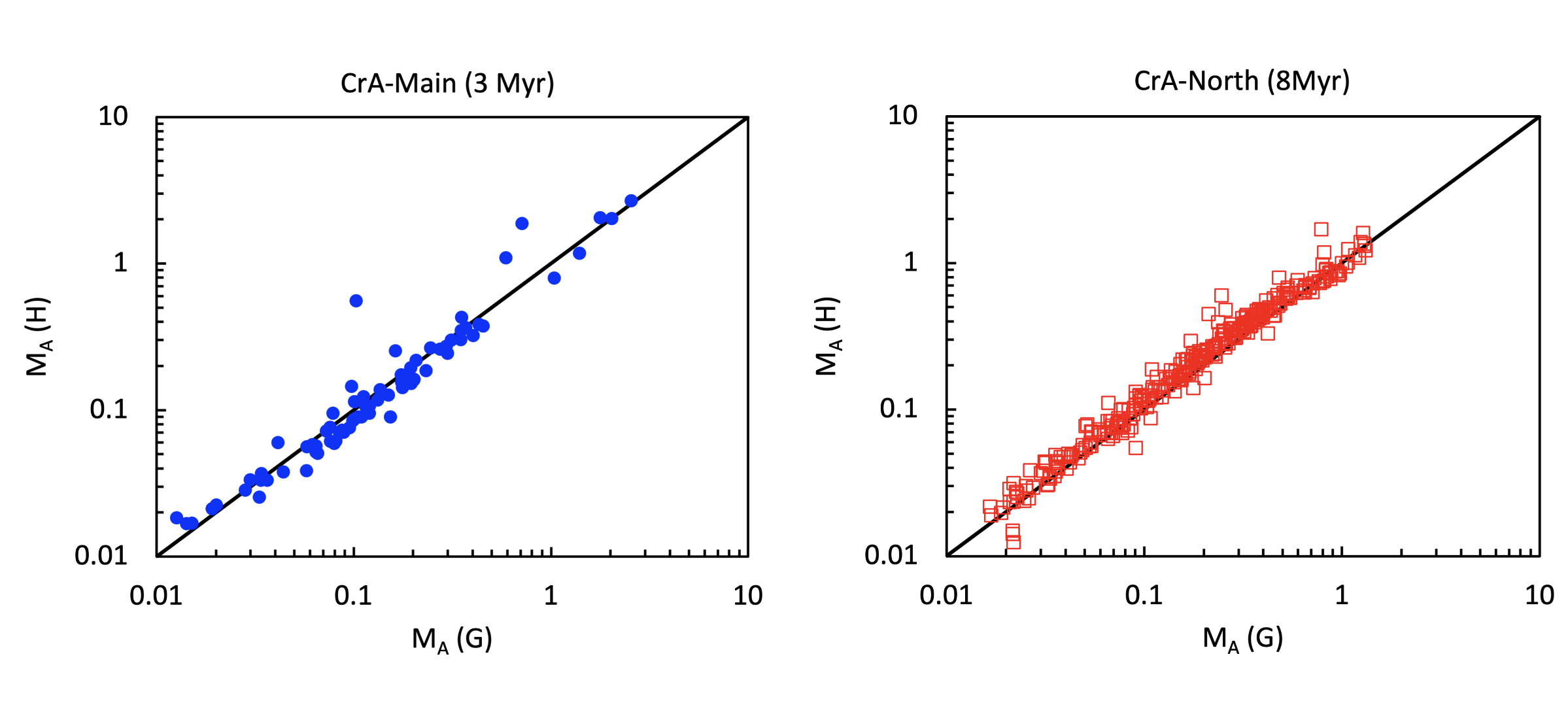}
      \caption{Comparison between masses for stars in the CrA complex derived from the $G$ - and $H$-band absolute magnitudes. The left panel is for CrA-Main and the right panel is for CrA-North. Solid lines mark identity.}
         \label{Fig:mass_bands}
\end{figure}

\begin{figure}[htp]
\centering 
  \begin{tabular}{ c c }
        \includegraphics[width=8.0cm]{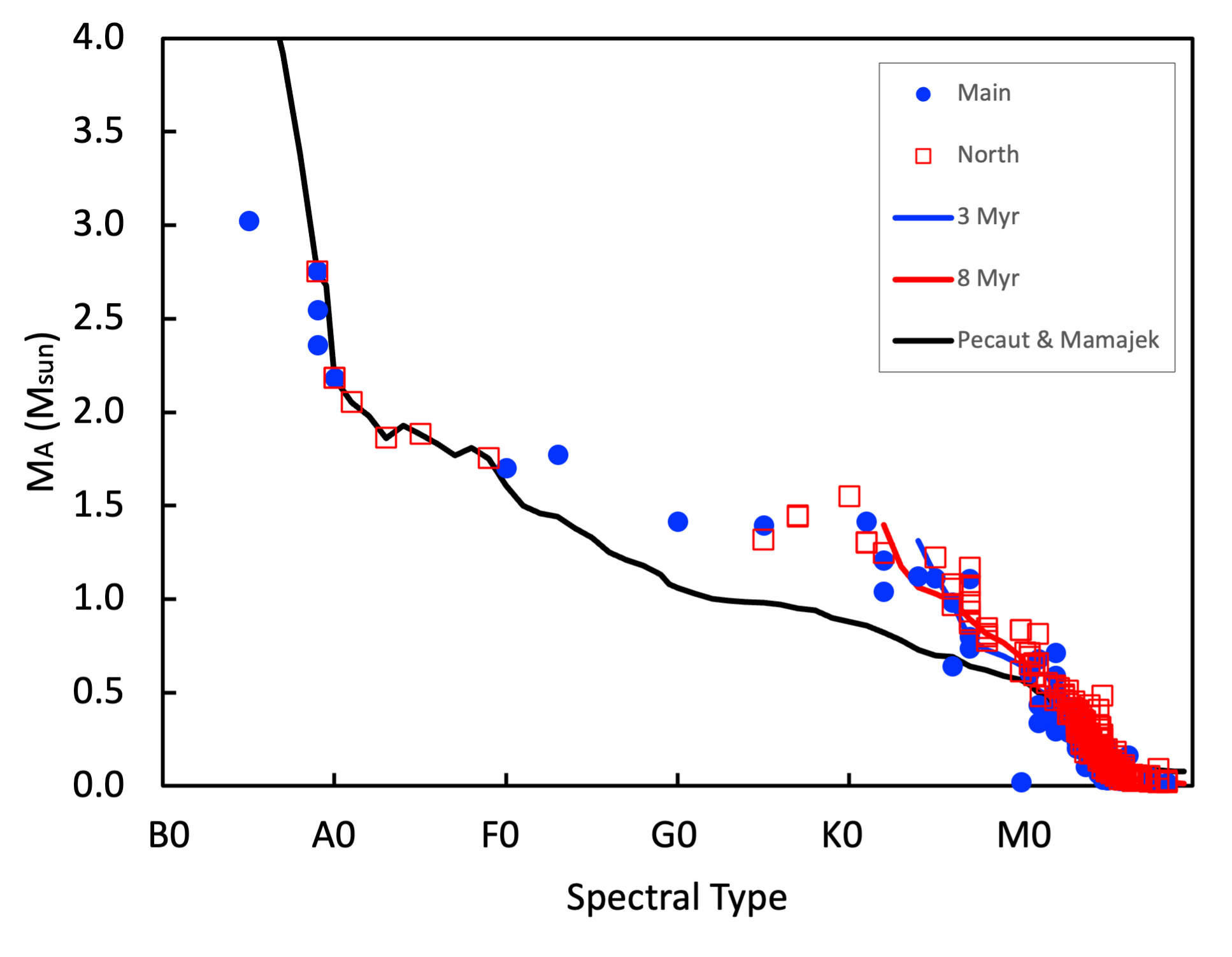} \\
        \includegraphics[width=8.0cm]{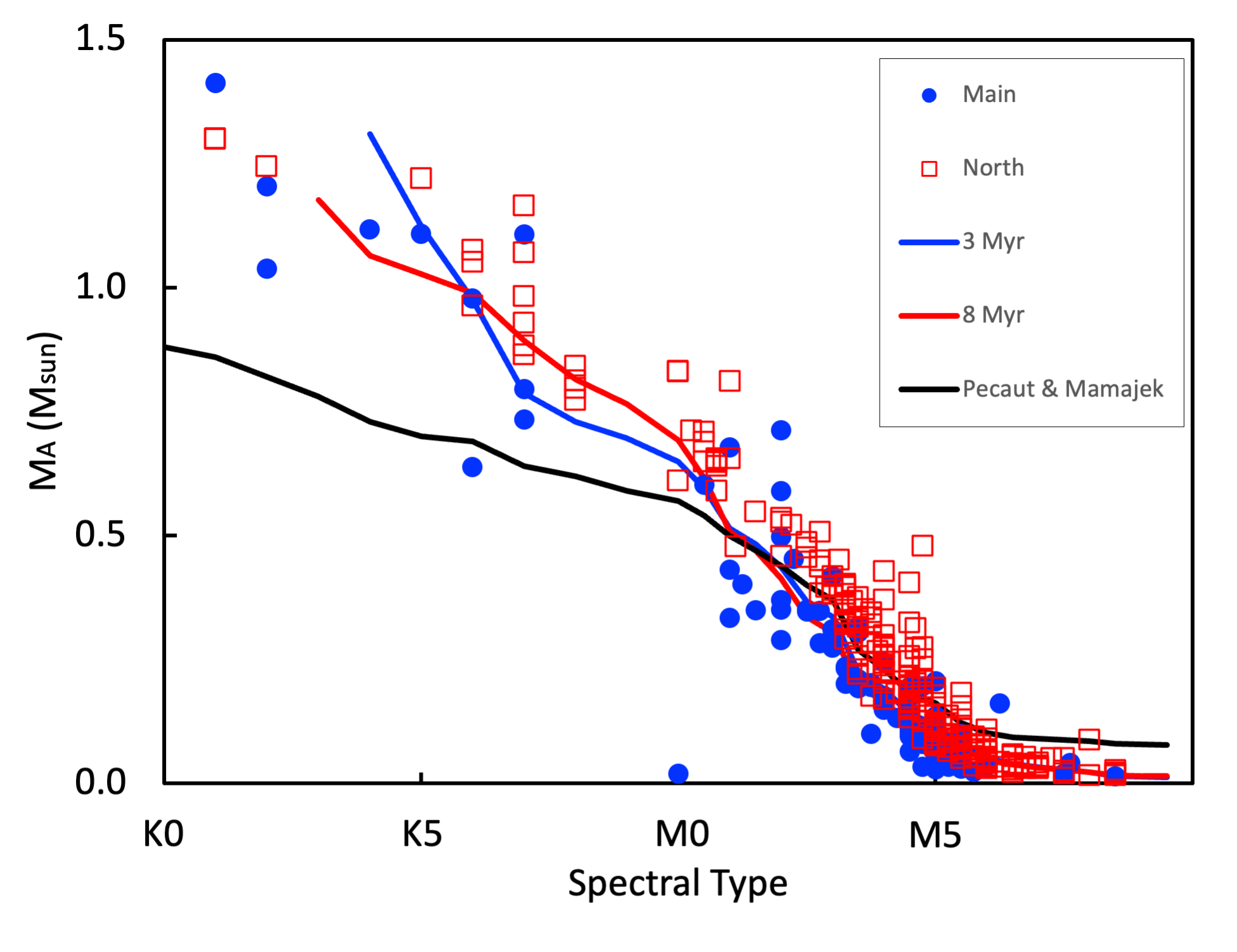} 
  \end{tabular}
\caption{Comparison between masses for stars in the CrA complex and spectral types (from \citealt{Esplin2022}). Filled blue symbols are for stars in CrA-Main; open red  symbols for those in CrA-North. The black line is the relation for old stars by \citet{Pecaut2013}; blue and red lines are the relations appropriate to \citet{Baraffe2015} isochrones of 3 and 8 Myr, respectively. The upper panel is for all stars; the lower panel is a blow-up on lower mass stars. }
         \label{Fig:mass_spectral_type}
\end{figure}

We completed a few consistency checks for the estimates of the masses. 
In figure~\ref{Fig:mass_bands}, we compare the masses obtained using Gaia $G$-band magnitudes with those obtained with $H$-band magnitudes adopting the \citet{Baraffe2015} isochrones, both for CrA-Main and CrA-North. The good fit that we find for both Main and North is a tell-tale sign that the temperatures of the stars are correctly reproduced by the isochrones. 
In figure \ref{Fig:mass_spectral_type} we plot the mass of the primary star as a function of the spectral type, together with the 3 and 8~Myr isochrones from \citet{Baraffe2015}, and the relations between mass and spectral type expected for main sequence stars from the table by \citet{Pecaut2013}. 
Since in the \citet{Baraffe2015} data spectral types were not given in the original tables, we considered the mass of stars that have temperatures corresponding to the spectral types in \citet{Pecaut2013}. The mass estimates are consistent with the young age of the stars in the CrA complex for stars more massive than about 0.3~\MSun, which corresponds to $M_H\sim 5$. This indicates that the radii of these stars are correctly reproduced by the isochrones. 
For masses lower than 0.3~\MSun, masses at a given spectral type are larger for stars in CrA-North than for stars in CrA-Main. These last agree fairly well with the \citet{Baraffe2015} models. This might indicate some failure in the models for the stars of lowest mass. We return to this issue in Section~\ref{subsection:ages}.

For many objects, the indication of the presence of companions comes from RUWE (49 objects with RUWE$>$1.4), or large RV variations (5 objects), or PMa (2 objects), or a combination of these techniques, for a total of 52 objects. The secondary of these stars is not imaged, and no period or semimajor axis is determined. Moreover, these binaries have different ranges of semimajor axes, depending on the technique used to detect them (see Figure~3 in \citealt{Gratton2024}). 
These plots show that binaries discovered through RV variations have a semimajor axis of at most a few au (generally $<$2~au), while binaries discovered through $RUWE$ have semimajor axes in the range 1-10 au, and those detected through PMa have a typical semimajor axis of around 10 au. Indeed, when considering astrometry, the semimajor axis for which maximum sensitivity is achieved depends on the length of the baseline considered: this is 34 months for the Gaia DR3 observations used in the case of $RUWE$, and 24.75~yr for the separation between the Hipparcos and Gaia DR3 epochs used for S/N(PMa). The different sensitivities of the various techniques indicate that we may better constrain masses and separations of binaries by combining different methods rather than considering only a single technique.

In Table \ref{tab:astrometric} we list objects that have indications for the presence of companions from PMa or RUWE. For these objects, we obtained solutions that are compatible with the observed values of RUWE and PMa, the scatter in RV, and the lack of detection in HCI (if available, else in Gaia), as done in \citet{Gratton2023a, Gratton2024, Gratton2025}. We did this by an exploration of the semimajor axis -- mass ratio plane using a Monte Carlo code. We adopted eccentric orbits, with uniform priors between 0 and 1 in eccentricity (which is in agreement with \citealt{Hwang2022} for this range of separations), 0 and 180 degrees in the angle of the ascending node $\Omega$, and 0 and 360 degrees in the periastron angle $\omega$, and left the inclination and phase to assume a random value. In addition, the period was used to fix the solution whenever it was available. Since in most cases only information from RUWE is available, uncertainties are large and only give order-of-magnitude estimates.

\begin{figure}[ht]
   \centering
   \includegraphics[width=8.0cm]{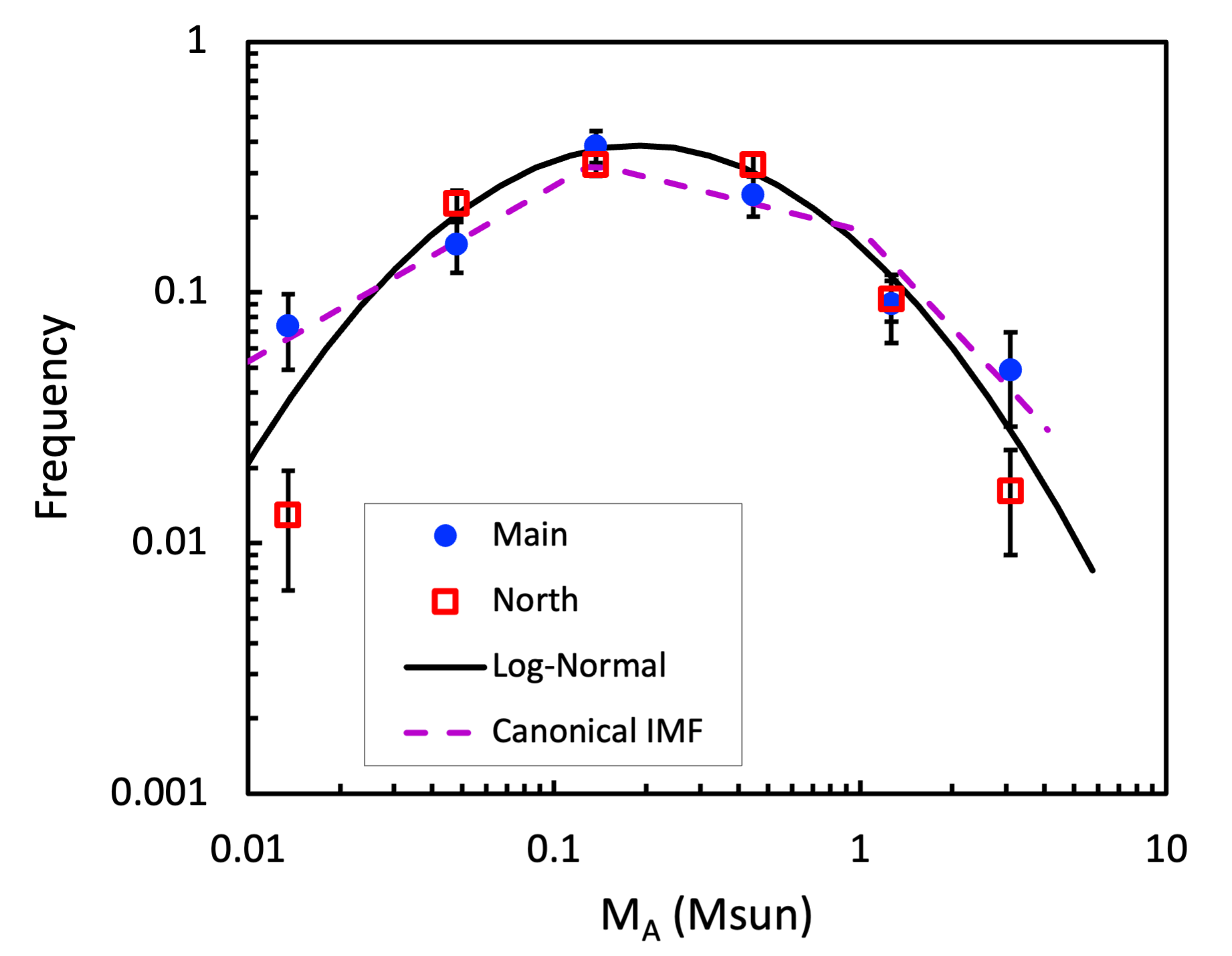}
      \caption{Mass function for primaries. CrA-Main  primaries are shown with filled blue circles and CrA-North primaries are shown with open red squares. The solid line is the best log-normal fit curve; the dashed line is the canonical initial mass function discussed by \citet{Kroupa2024}. }
         \label{Fig:mass_function}
\end{figure}

As a final check of the masses adopted for our targets, Figure \ref{Fig:mass_function} shows the mass function we obtained for the primaries for CrA-Main and CrA-North. The same log-normal curve (peak value 0.19 \MSun, standard deviation 0.53 \MSun) fits both groups. This mass function is similar to the predictions of star formation in turbulent media by \citet{Haugbolle2018} and to the initial canonical broken power law mass function described by \citet{Kroupa2024}. This suggests that the samples we are considering are complete for stars and brown dwarfs with mass $>0.02$~\MSun\, in both CrA-Main and CrA-North, while the list is likely incomplete for the brown dwarfs of lower mass in CrA-North. This can be explained by their older age and faintness. These brown dwarfs were not considered in the fit.

\subsubsection{Completeness of companion detections}

\begin{figure}[ht]
   \centering
      \includegraphics[width=8.0cm]{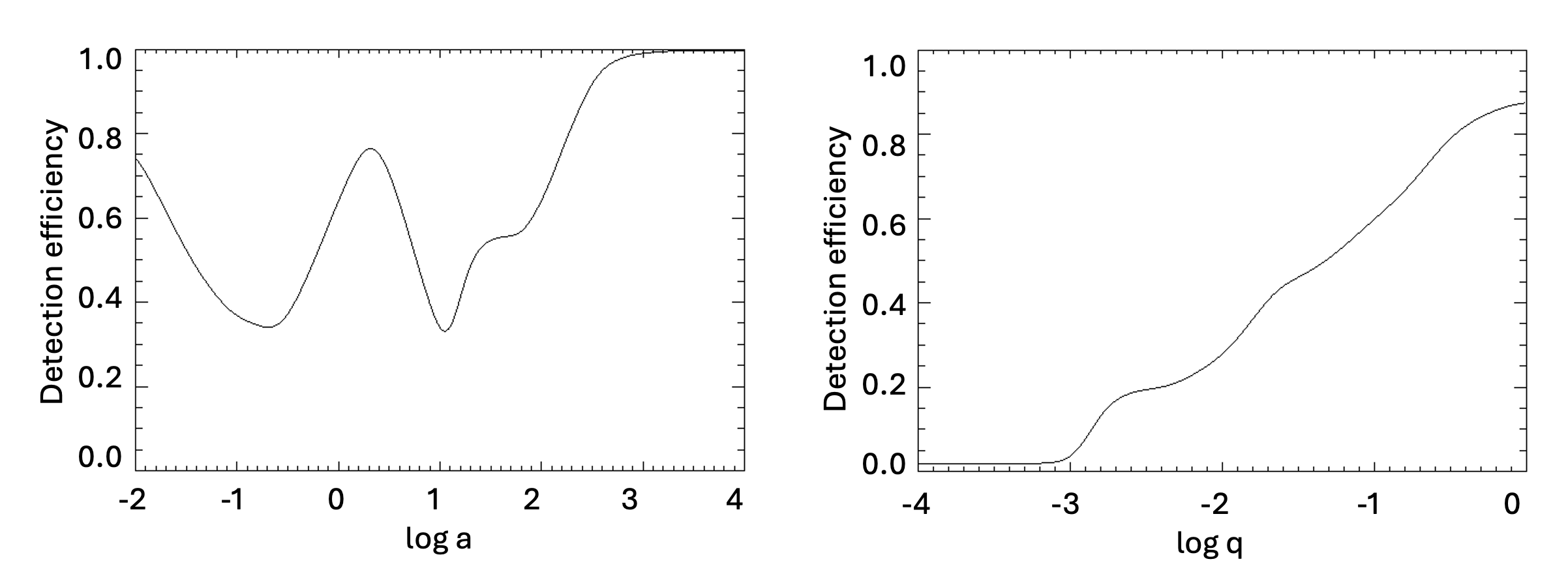}
      \caption{Left panel: Detection efficiency of stellar and massive brown dwarf companions ($\log{q}>-1.5$) for primaries more massive than 0.8 \MSun as a function of semimajor axis, $a$. Right panel: Detection efficiency for all companions of primaries more massive than 0.8 \MSun as a function of the mass ratio, $q$ }
         \label{Fig:efficiency}
\end{figure}

We found a total of 130 companions to the 434 stars in the CrA-complex. The frequency of companions detected is much higher if we only consider massive stars ($>0.8$ \MSun: 38 out of 51 stars). The real frequency should be higher because our search is incomplete. We estimated the completeness of companion detections for stars with primary mass $M_A>0.8$~\MSun using the same Monte Carlo approach described in \citet{Gratton2023a, Gratton2024, Gratton2025} where the reader may find a full description of the procedure. Briefly, in order to have enough statistics for this rather small sample of stars (a total of 51 systems), we considered 300,000 extractions for each of the stars in our sample (with appropriate values of the magnitude, mass, parallax and age) with random values of the semimajor axis $a$, mass ratio $q$, inclination $i$ and phase, assuming circular orbits (this choice is commented on in \citealt{Gratton2023a}). A total of 15.3 million companions were simulated. We considered uniform distributions in the logarithm for $a$ and $q$ (with $0.01<a<10000$ and $0.0001<q<1$), uniform distributions of phase between 0 and 1, and isotropic distributions of inclinations. For each target and methodology (visual, eclipsing, spectroscopic, and astrometric binaries), we considered whether the appropriate datasets were available for the star considered and then determined the signal expected for each simulated companion according to the various methods considered (transit/eclipses, RVs, Gaia imaging/HCI, RUWE, PMa). When the data are available for the star considered, we compared this signal with the detection limits appropriate for each method. We considered that each simulated companion that produces a signal above the detection limits is detected. In order to reduce local fluctuations due to low number statistics, we finally smoothed the detection distribution in the ($\log{a}$ and $\log{q}$) plane with a Gaussian kernel of 0.1 dex; this ensures that of the order of 6,000 simulated planets are considered for each position in the ($\log{a}$ and $\log{q}$) region explored by this simulation.

Figure \ref{Fig:efficiency} illustrates the completeness of the companion search as a function of the semimajor axis $a$ and of the mass ratio $q$. Search completeness is high for wide companions (mainly due to detections from Gaia and NIR photometry) and has a secondary peak at about a few au (due to detections based on the RUWE parameter). However, our search is probably missing about half of the close companions and of the companions in the separation range 10-200 au. 
Not surprisingly, most of the companions that are missed are low mass.

\subsubsection{Overall binary properties}
\label{subsection:overall}

\begin{figure}
   \centering
   \includegraphics[width=8.0cm]{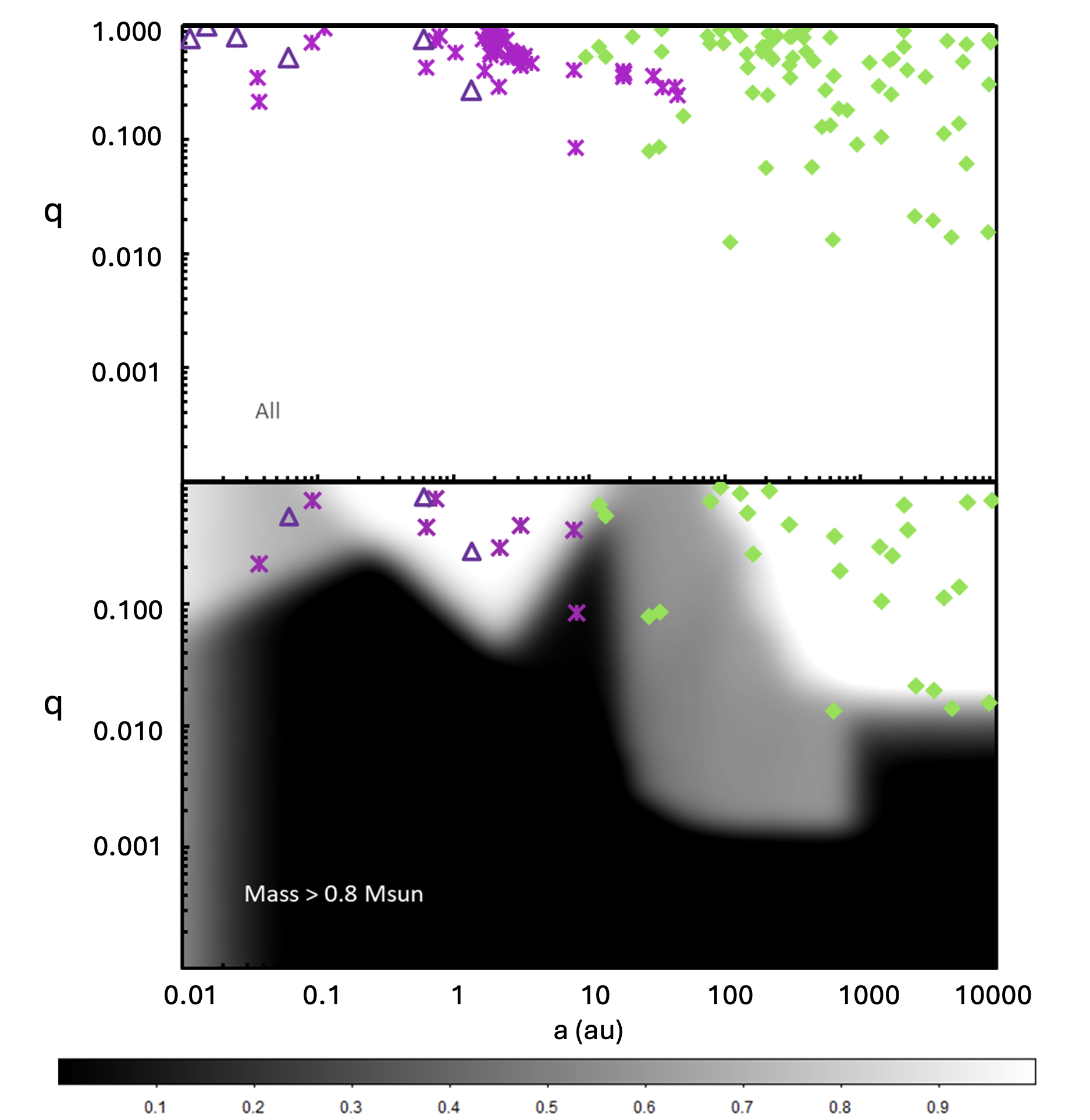}
        \caption{Distribution of companions to the star from our sample in the semimajor axis ($a$) versus mass ratio ($q$) plane. Green diamonds  are companions detected in imaging, magenta asterisks are astrometric binaries, and purple open triangles are spectroscopic and eclipsing binaries. The top panel is for all stars and the bottom panel is for systems with primaries more massive than 0.8 \MSun. In the bottom panel, the gray-scale coverage shows the map of the efficiency of binary retrieval using the combination of all techniques considered in this paper, for systems with primaries that have a mass $>0.8$ \MSun. The gray-scale goes from 0 to 1 as shown by the bar at the very bottom of the figure. }
         \label{Fig:binaries}
\end{figure}

\begin{figure}
   \centering
   \includegraphics[width=8.0cm]{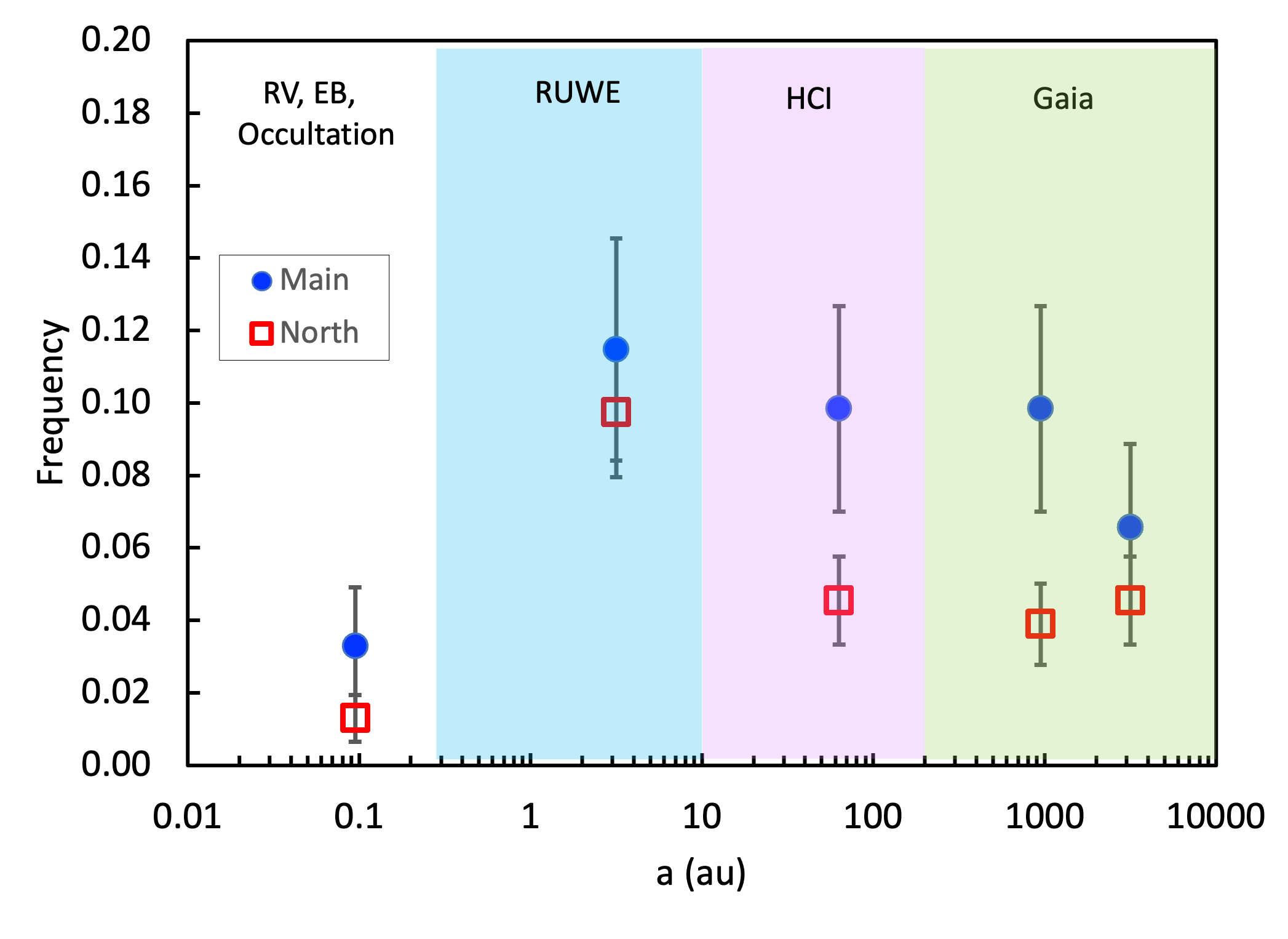}
      \caption{Observed frequency of companions in different bins of semimajor axis $a$ for stars in CrA-Main (filled blue circles) and CrA-North (open red squares). The different bins correspond to different discovery methods: $<0.3$ au: radial velocities (RV), eclipsing binaries (EB), disk occultation; $0.3<a<10$ au: RUWE; $10-200$ au: high contrast imaging (HCI); $>200$ au: Gaia imaging.}
         \label{Fig:frequency}
\end{figure}

Table \ref{tab:binaries} summarizes the statistics of companions from our search: we provide the number of stars for which the relevant data are available and the number of detections using the various techniques. In Appendix~\ref{AppendixB:data} we also report the same table (Tab.\ref{tab:binaries_techniques}) only for systems with mass of the primary $>0.8$~\MSun. The latter sample is consistent with those considered in \citet{Gratton2023b, Gratton2024, Gratton2025}. Figure \ref{Fig:binaries} shows the distribution of the mass ratios $q$ as a function of the semimajor axis $a$ for the detected companions. In the bottom panel of this figure we show in the background the map of the efficiency of binary retrieval using the combination of all techniques considered in this paper, for systems with primaries having a mass $>0.8$ \MSun. By comparing this figure with similar ones presented in \citet{Gratton2023b, Gratton2024, Gratton2025}, it is clear that all detected companions are in the upper part of the diagram that corresponds to stellar/brown dwarf companions. This is due to the lack of sensitivity to planetary companions.

\begin{table}
\centering
\caption{Binary statistics.  }
\begin{tabular}{lcc}
\hline 
 & CrA-Main & CrA-North \\
 \hline
Total & 122 & 308 \\
\hline
\multicolumn{3}{c}{Technique}\\
\hline
Gaia & 109 (0.90) & 308 (1.0) \\
HCI & 32 (0.26) & 15 (0.05) \\
RUWE & 99 (0.81) & 308 (1.0) \\
RV & 12 (0.10) & 43 (0.14) \\
\hline
\multicolumn{3}{c}{Detection method}\\
\hline
Visual & 34 (0.28) & 37 (0.12) \\ 
RV/EB & 3 (0.02) & 3 (0.01) \\
Astrometry & 16 (0.13) & 37 (0.12) \\
\hline
\end{tabular}
\tablefoot{In the "Technique" subtable, we report the number of stars observed with the different techniques (in parenthesis the frequency with respect to the total sample). In the "Detection method" subtable, we report the number of stars detected with the different methods (in parenthesis the frequency with respect to the total sample). }
\label{tab:binaries}
\end{table}

We compared in Figure \ref{Fig:frequency} the observed frequency of companions in CrA-Main and CrA-North considering various bins of semimajor axis $a$. To have a wider statistics, we considered here all stars irrespective of their mass because the mass functions of the two populations are similar (see Section \ref{subsubsection:parameters}). We only considered companions with mass $>0.02$~\MSun because 8~Myr old companions of lower mass are too faint to be detected by Gaia. The frequency of companions in CrA-Main is higher than in CrA-North in all bins. However, a more careful exam reveals interesting facts. Very few close companions ($a<0.3$) are detected, so the first bin is scarcely significant. The difference between Cra-Main and CrA-North is not significant for the range 0.3-10 au, where detection is mainly due to high values of the RUWE parameter, which is available for a large fraction of the stars in both samples, and then a similar level of completeness is achieved. The higher fraction of companions in CrA-Main in the range of separation 10-200 au is consistent with the much larger fraction of stars that were observed using high-contrast imaging in this region. Relevant is the difference for wide visual companions ($200<a<3000$ au, mainly detected using Gaia) that varies from $\sim$0.08 for CrA-Main to $\sim$0.03 for CrA-North. The excess of companions in Cra-Main is significant and is likely physical, because the search completeness should be very similar for CrA-Main and CrA-North in this range of separation, and because very few chance alignments are expected for these moderate separation binaries. This is also confirmed by the lower and not significant excess at even wider separation ($a>3000$ au). This result may suggest that secondaries with semimajor axis of about 1000 au are frequently lost on a timescale of a few millions of years in the CrA complex. Alternatively, the original density of CrA-North was higher than the present density of CrA-Main, making the formation of wide binaries more difficult (see, however, Section \ref{subsection:disks} for arguments against this hypothesis).

\subsection{Ages of CrA-Main and CrA-North}
\label{subsection:ages}

\begin{figure*}
   \centering
   \begin{tabular}{ c c }
     \includegraphics[width=8.0cm]{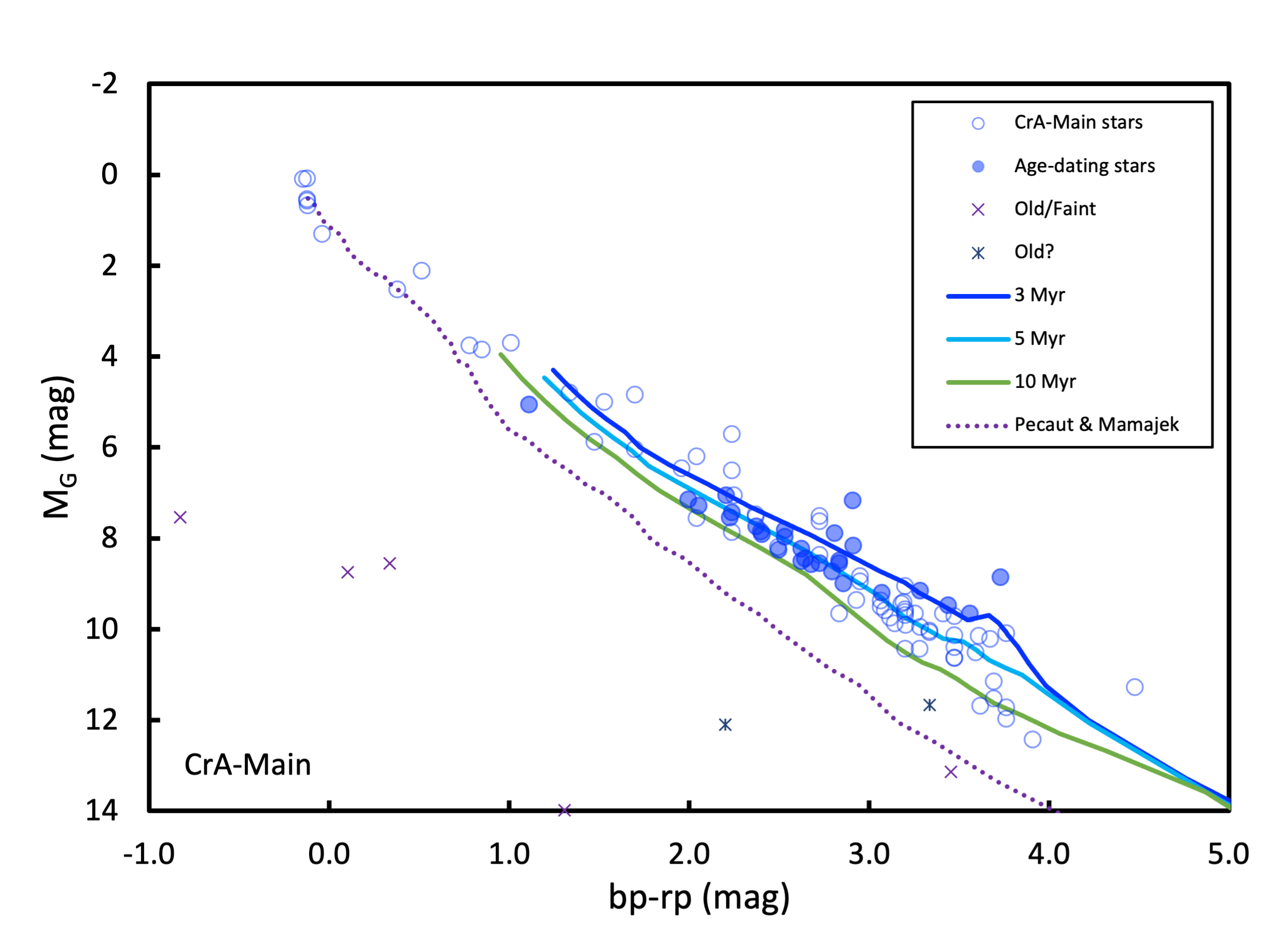} &
     \includegraphics[width=8.0cm]{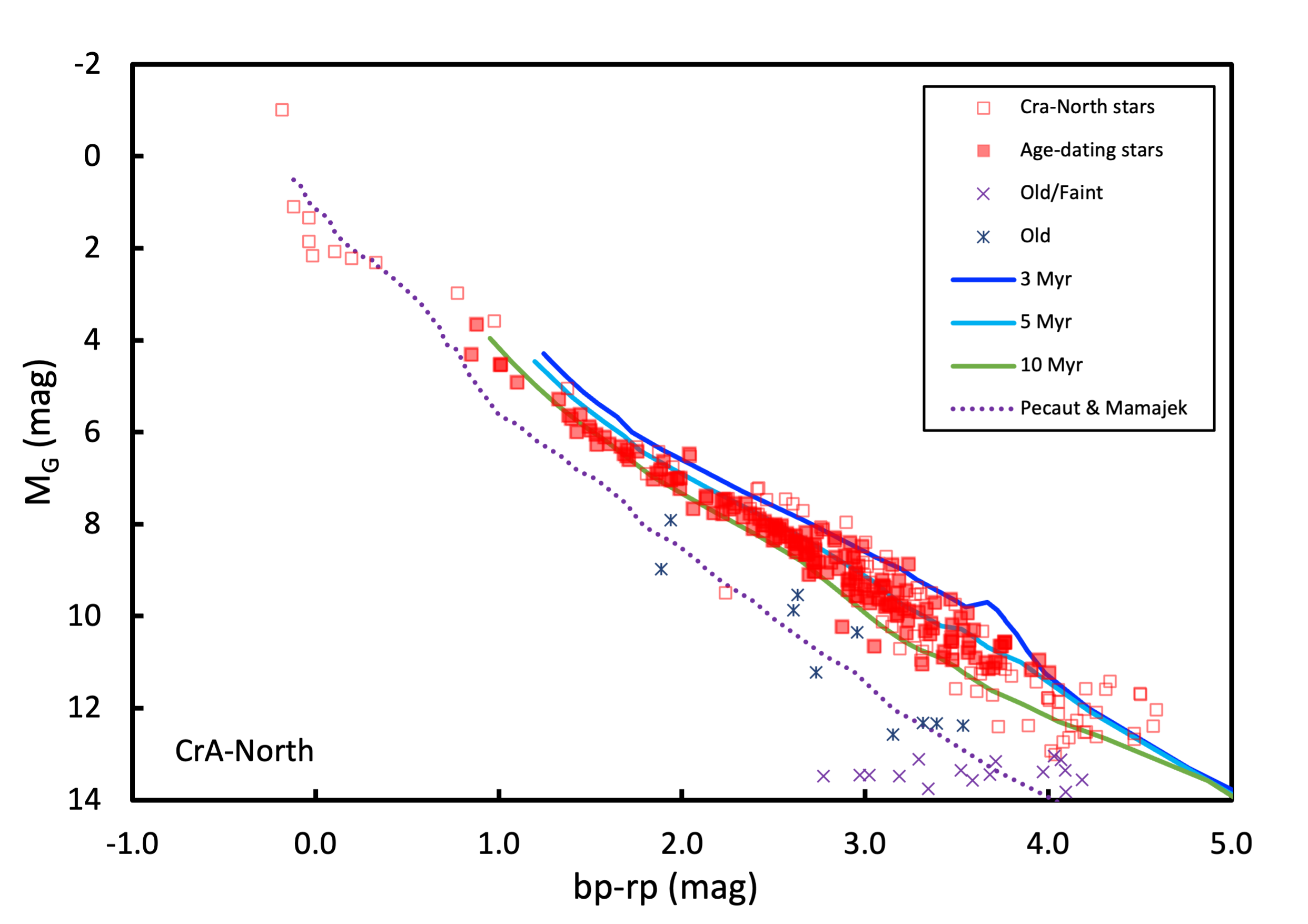} \\
     \includegraphics[width=8.0cm]{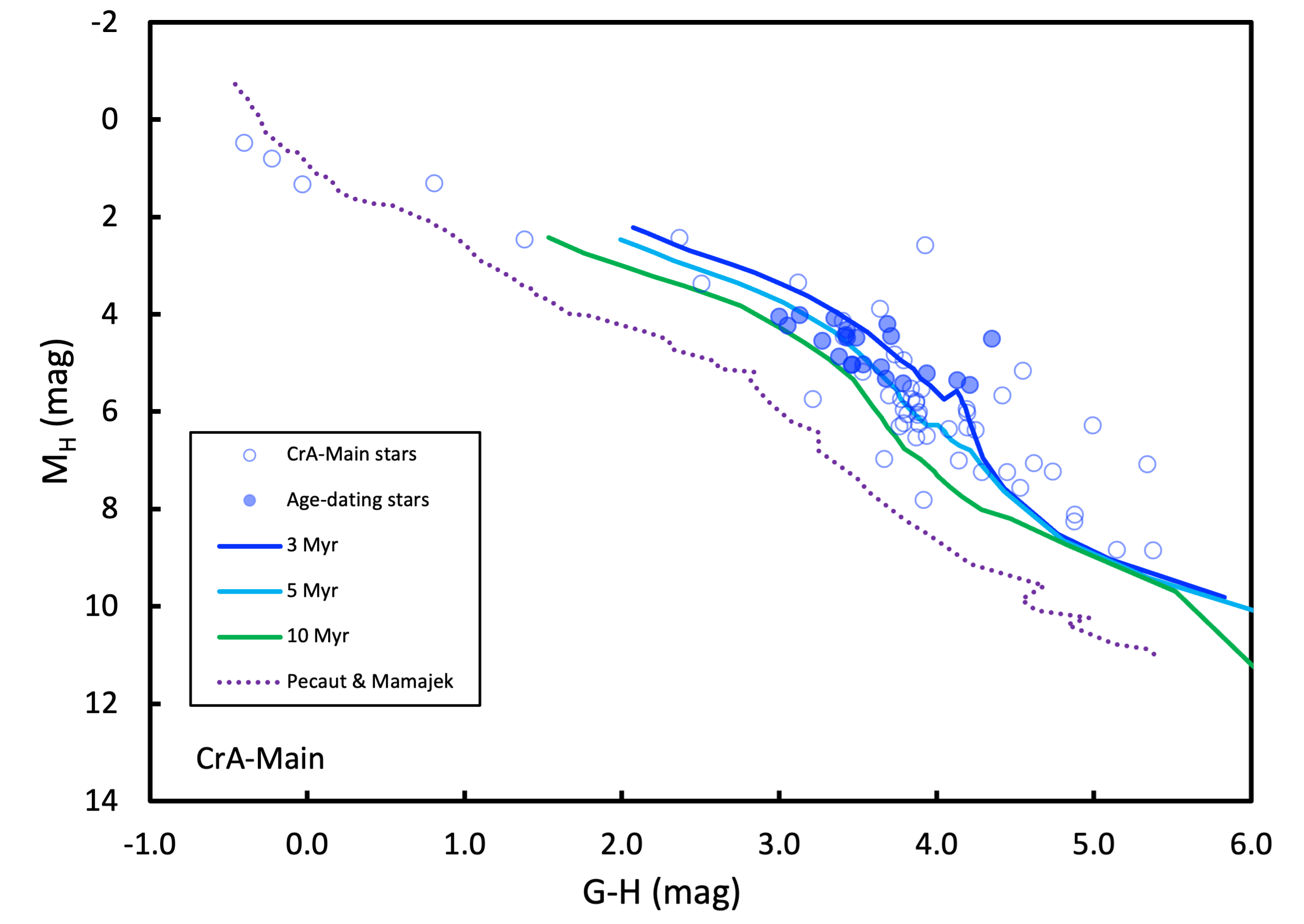} &
     \includegraphics[width=8.0cm]{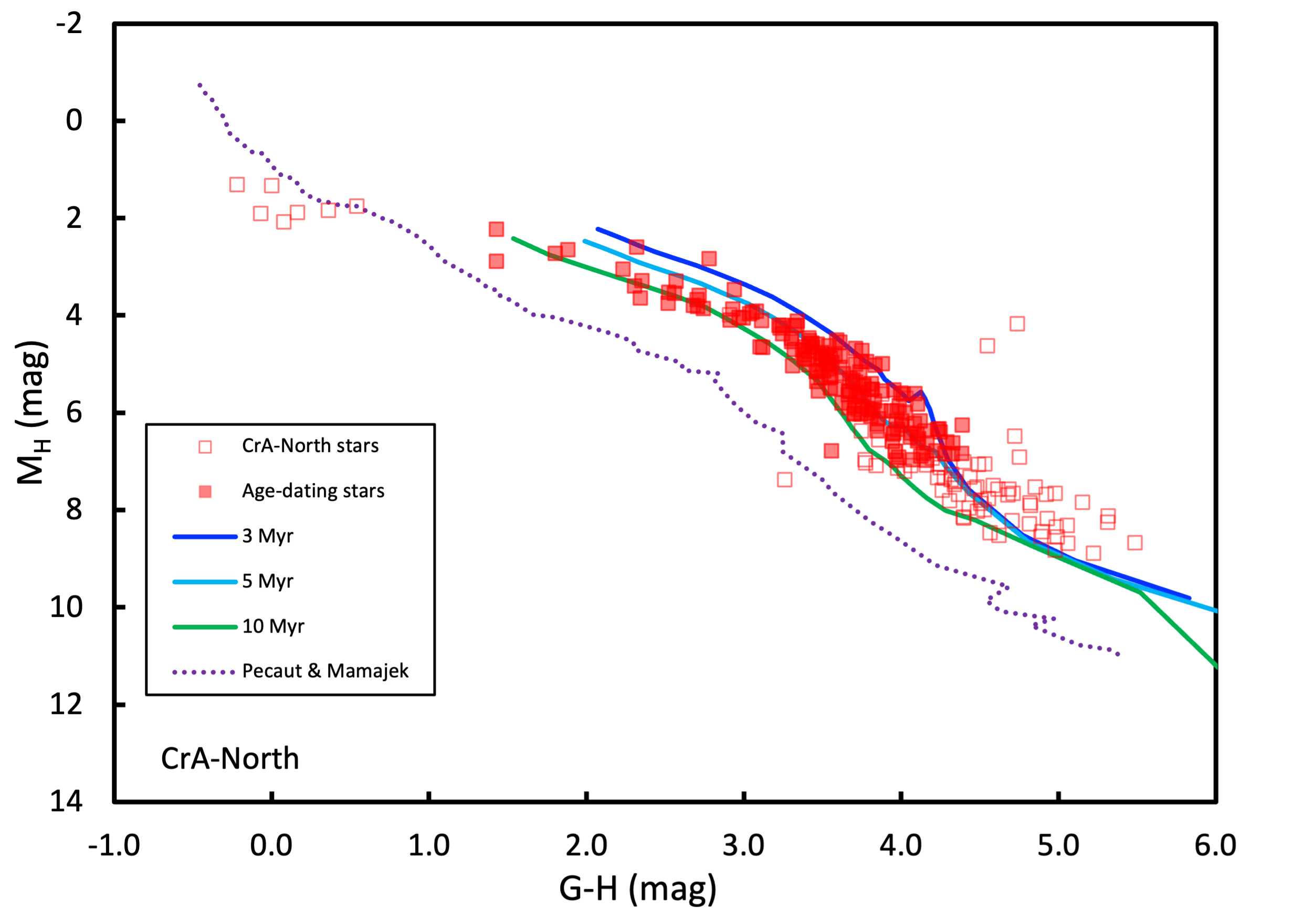}
  \end{tabular}
  \caption{Color magnitude diagrams for stars in the CrA-complex. The top panels show the ($G$, $BP-RP$) magnitudes and colors for Main and North, respectively. The bottom panels show the ($H$, $G-H$) magnitudes and colors. Filled symbols are for stars used for the age-dating of the CrA-Main and North groups, open symbols are all the other stars. Overimposed are the isochrones from \citet{Baraffe2015} for different ages, and the main sequence by \citet{Pecaut2013}, which is representative of an old population. We marked with purple crosses the stars with apparent G-magnitude $>$20, which we define as Old/Faint stars, and with blue asterisks the stars that lie closer to the main sequence from \citet{Pecaut2013} than to the 10~Myr isochrone from \citet{Baraffe2015}, identified as Old.}
         \label{Fig:cmd}
\end{figure*}
A crucial aspect of this study concerns the age of the stars in the CrA-complex. The age of very young stars can be determined by a comparison of their positions in the CMD with theoretical isochrones. 
In Figure~\ref{Fig:cmd} we show all the stars in CrA-Main and CrA-North for which the $G$-band magnitude and the $(bp-rp)$ color are available (top panels), and the $H$-band magnitude and the $(G-H)$ color are available (bottom panels). We identify as "Old/Faint" objects the ones with an apparent $G$-band magnitude higher than 20~mag, and as "Old" objects the ones that in the CMD are closer to the main sequence (as measured by \citealt{Pecaut2013}) than to the 10~Myr isochrone from \citet{Baraffe2015}. 
Before providing an age estimate for all the stars in the CMD we have taken into account different aspects: an appropriate consideration of stellar multiplicity, the presence of disks or accretion features, interstellar absorption, and the accuracy of isochrones in reproducing colors of the stars. In order to conservatively take into account all these aspects, we only considered stars that appear to be single or with only wide companions, which does not affect the photometry. The choice of considering the $(G, BP-RP)$ and $(H, G-H)$ CMD enables us to avoid the wavelengths that are most affected by accretion at short wavelengths and contamination by disks at long ones. We limited the study of the ages to stars with $M_H>3$ because the ages are uncertain for massive stars that rapidly reach the zero age main sequence, and to stars with $M_H<5.5$ because the \citet{Baraffe2015} isochrones yield low ages for low-mass stars \citep{Squicciarini2021}, possibly because of inappropriate consideration of stellar activity \citep{Feiden2016} (see also Section \ref{subsubsection:parameters}). Moreover, we only considered stars with masses $\leq 1.4$ \MSun. With all these limitations, we could derive the ages for 29 stars in CrA-Main and 94 stars in CrA-North. This represents $\sim$24\% and 30\% of the stars, respectively, and they are indicated as filled symbols in Figure~\ref{Fig:cmd}.

Considering these stellar subsamples, we find that the linear average age over all these stars for CrA-Main is $5.1\pm 0.5$ Myr, with a standard deviation of 2.8~Myr, and $6.7\pm 0.3$ Myr, with a standard deviation of 2.3~Myr for CrA-North. The uncertainties are purely statistical and do not include systematics. Note that the average of logarithms is possibly more appropriate for CrA-Main considering how isochrones scale and the range of ages in this group; this provides an average value of $4.1\pm 0.7$ Myr.
These ages are younger or in agreement than the values obtained by \citet{Ratzenbock2023b} ($8.5^{+2.0}_{-2.4}$ and $11.6^{+0.5}_{-0.8}$ Myr using the PARSEC isochrones, \citealt{Costa2025}, and $6.1^{+2.4}_{-0.6}$ and $6.6^{+0.3}_{-0.8}$ using the \citet{Baraffe2015} isochrones, for CrA-Main and CrA-North, respectively), but older than the estimate of $<3$~Myr by \citet{Sicilia-Aguilar2013}. They are very similar to the values obtained by \citet{Galli2020}. We notice that even in CrA-Main, the youngest stars have ages of about 2 Myr. However, IR observations of CrA-Main clearly show that star formation is still ongoing in this part of the CrA-complex. This might be because stars with ages $<2$ Myr in CrA-Main are still heavily embedded in the CrA-cloud. Strong absorption explains why there is no counterpart in Gaia for about 30\% of the CrA-Main members detected only in Spitzer and Herschel images \citep{Peterson2011, Sicilia-Aguilar2013}. Nine stars are deeply embedded in the mass range considered for age derivation and without a close companion, they would be considered for age determination of the CrA-Main group if they were not embedded. To have a less biased estimate of the age of CrA-Main, we have then attributed an age of 1 Myr to all of them. The average age of CrA-Main in this case is $2.9\pm 0.5$ Myr. We consequently adopt a mean age of $3\pm 1$ Myr for this group of stars. This matches one of the grids used by \citet{Baraffe2015}. For CrA-North, the closest match in the grid of \citet{Baraffe2015} is 8 Myr. We then used these isochrones in this paper. 

\begin{figure}[ht]
   \centering
   \includegraphics[width=8.0cm]{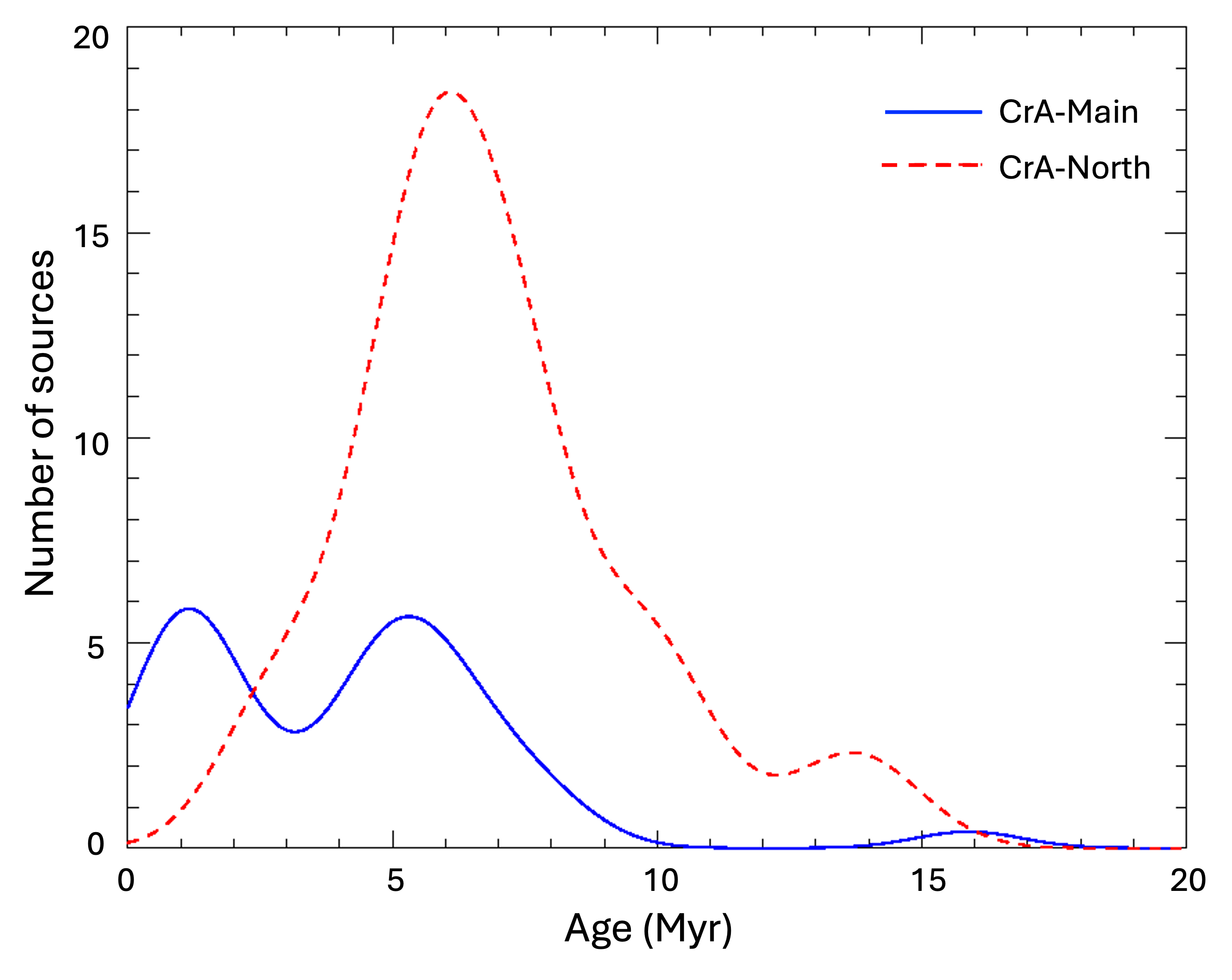}
      \caption{Generalized histogram of the age distribution for the stars belonging to CrA-Main (blue solid line) and CrA-North (red dashed line) for which we have an age estimation. A gaussian kernel of 1~Myr has been used to generate this plot. }
         \label{Fig:histo_age}
\end{figure}

In Figure~\ref{Fig:histo_age} we show a generalized histogram of the distribution of the ages of the stars in CrA-Main and CrA-North for those stars with age estimates. To generate this plot, we used a Gaussian kernel with a standard deviation equal to 1 Myr. 
We include the nine single stars in CrA-Main that are in the mass range appropriate for the age derivation but are strongly absorbed and then miss adequate Gaia photometry, assuming a 1 Myr age. 
We notice that the distribution of ages in CrA-Main looks bimodal, with a peak with age between 4--7~Myr, and a second peak with age $<3$ Myr. This might suggest either that star formation occurs in episodes, separated by rather quiescent phases, or that there might be a contamination of stars belonging to CrA-North that are erroneously associated to CrA-Main, creating de facto the peak at older ages. Conservatively, we may conclude that the members of the CrA-Main have a rather wide distribution in ages, ranging from about 7 Myr to very low values. We finally notice that the distribution of ages of CrA-North also shows a secondary peak at about 14 Myr, similar to the age of the majority of stars in Sco-Cen \citep{Ratzenbock2023b}. 

\subsection{Disk properties}
\label{section:disk}

Another precious piece of information comes from the circumstellar disks observed in the CrA-complex. To identify objects with disks, we use the slope of the spectral energy distribution (SED) between the WISE bands $W1$ (at 4.6 $\mu$m) and $W3$ (at 11.6 $\mu$m), following the widely used classification of Lada \citep{Lada1987, Lada2006}. This is based on the parameter $\alpha,$ that is 
\begin{equation}
    \alpha = \frac{d\log{\lambda F_{\lambda}}}{d\log{\lambda}}
,\end{equation}
where $F_{\lambda}$ is the flux density at the given $W1$ and $W3$ wavelength. According to this definition, sources are classified into five categories, as reported in Table~\ref{Tab:disk_bearing_classification}. In the first category are objects where the naked photosphere of the star is observed, meaning diskless objects. Anemic disks are defined as disks that are homogeneously depleted throughout their radius, where the disk flattens uniformly. The thick disks show a typical class II disk. Flat spectrum disks refer to thick class I resembling disks, and the last category includes protostars that are still embedded in their forming nebula. 
Moreover, we introduce the value $\alpha^{\prime}$ for the pure scope of plotting a logarithmic scale on the x-axis in some figures. $\alpha^{\prime}$ is defined as $\alpha$+3.896, where the latter value is the zero-point magnitude value. 

\begin{table}[!ht]
\caption{Definition of disk-bearing objects according to the $\alpha$ value.}. 
    \centering
    \begin{tabular}{cc}
    \hline
       Source classification  & $\alpha$ value \\
    \hline   
    Photosphere  & $\alpha <$-2.3 \\
    Anemic disk  & -2.3$< \alpha <$-1.8 \\
    Thick disk & -1.8$< \alpha <$-0.3 \\
    Flat spectrum & -0.3$< \alpha <$0.3 \\
    Protostar & $\alpha >$0.3 \\
    \hline
    \end{tabular}
\tablefoot{In Figures~\ref{Fig:alpha_age} and ~\ref{Fig:alpha_ag}, we use the $\alpha^{\prime}$ value, which is defined as $\alpha$+3.896. The latter value is the zero-point magnitude value. }
    \label{Tab:disk_bearing_classification}
\end{table}

In general, there is a very good correlation between the classification obtained using the $\alpha$ index and that of \citet{Esplin2022} (see Table \ref{tab:disk_esplin}), which is also based on the SED. 
There are in total 308 stars that have both data. As shown in Table~\ref{tab:disk_esplin}, the naked photospheres correspond to class III of \citet{Esplin2022}, which also includes the cases of anemic disks, with some cases classified as debris. The full disks and flat/I in \citet{Esplin2022} correspond mainly to the thick disks and protostars in \citet{Lada1987} and \citet{Lada2006}.

\begin{table}[!ht]
\caption{Comparison between disk classification of this paper (rows) and from \citet{Esplin2022} (columns).}
\centering
\begin{tabular}{lcccccc}
\hline 
Type & III & debris & evol. & transit. & full & flat/I \\
\hline
Photosphere &192 & 21 & 2 & 0 & 2 & 0 \\
Anemic      & 16 &  7 & 3 & 0 & 1 & 0 \\
Thick       &  1 &  0 & 5 & 2 &48 & 0 \\
Flat        &  0 &  0 & 0 & 0 & 3 & 1 \\
Protostars  &  0 &  0 & 0 & 0 & 0 & 4 \\
\hline
\end{tabular}
\label{tab:disk_esplin}
\end{table}

\begin{figure}
   \centering
   \includegraphics[width=8.0cm]{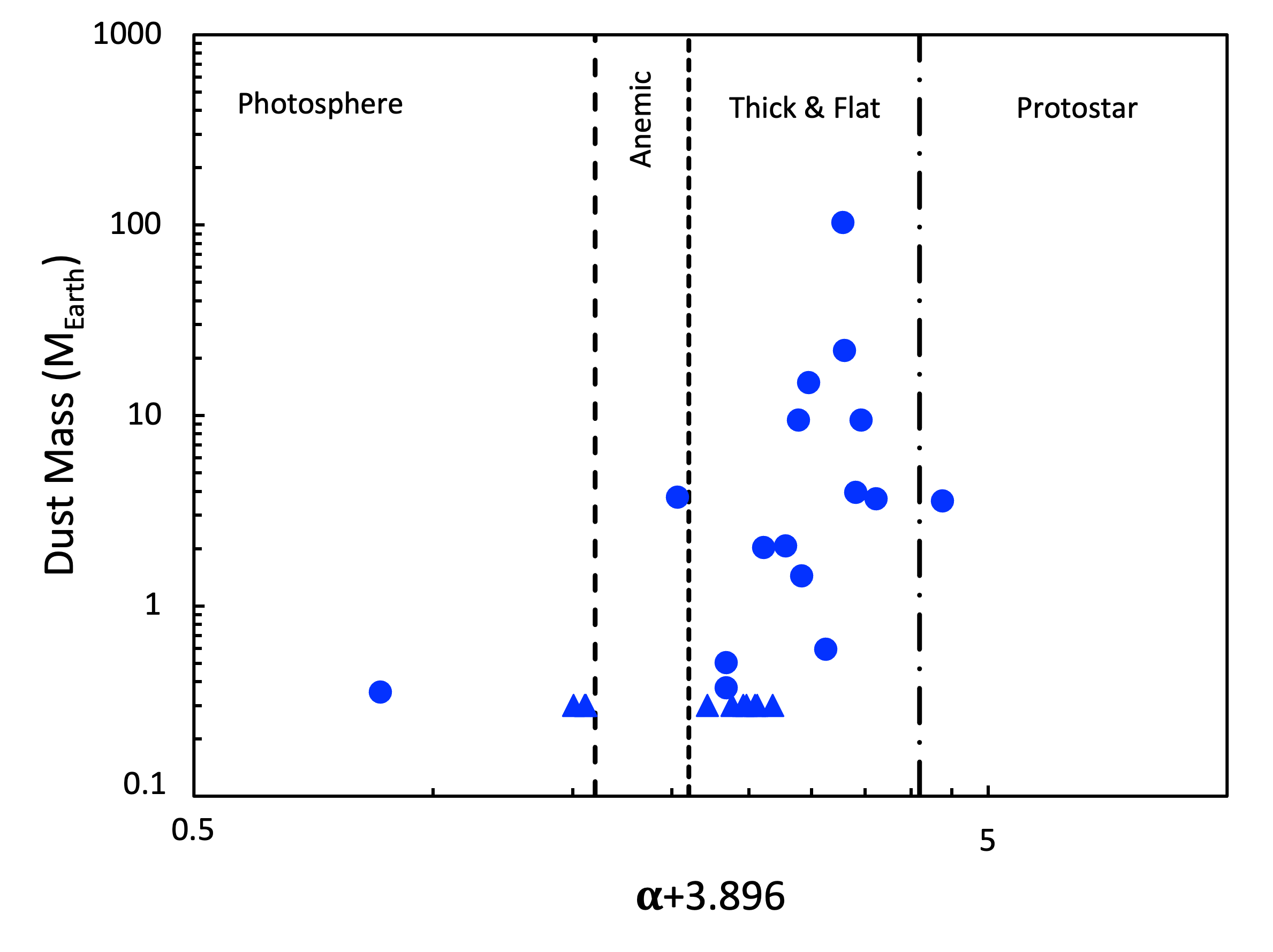}
      \caption{Correlation between the $\alpha^{\prime}$ value and the dust mass in the disks observed with ALMA by \citet{Cazzoletti2019}. Blues circles are actual measures with ALMA, triangles are upper limits. The dashed lines separate the different categories of disks as labeled in Table~\ref{Tab:disk_bearing_classification}. Thick disks and flat spectra are merged together}
         \label{Fig:alpha_dustmass}
\end{figure}

The $\alpha$ index is correlated with other disk quantities, such as the dust mass derived from sub-mm observations. Figure \ref{Fig:alpha_dustmass} shows how the values of $\alpha^{\prime}$ correlate with the dust masses derived by \citet{Cazzoletti2019} using ALMA data. In this plot, we assumed that non-detections correspond to an upper limit of 0.3 $M_{\rm Earth}$.

\section{Discussion}
\label{section:discussion}

\subsection{CrA-Main as an infant cluster}
\label{subsection:CrA-Main_infant}

In Appendix~\ref{Appendix_c} we report the different estimates of the mass budget of the CrA-complex. The total mass in stars in CrA-Main, obtained by the sum of the stars analyzed in this paper, is 69~\MSun. 
Comparing this value to the total mass of the CrA-Cloud in the direction of CrA-Main (see Appendix~\ref{Appendix_c} and reference therein, we obtain that between 5\% (if M$_{\rm{TOT \, Main}}$=950~\MSun is assumed, \citealt{Alves2014}) and 7\% (if M$_{\rm{TOT \, Main}}$=1330~\MSun is assumed, \citealt{Cambresy1999}) of the current cloud mass is in stars.

We also notice that half of the stars in CrA-Main are projected within a distance of 0.75 pc from R CrA itself, which roughly represents the center of CrA-Main. The projected stellar half-mass radius of CrA-Main is 0.43 pc, which is lower than the half-count radius of 0.75 pc. In turn, the latter value agrees with the estimate of 0.7 pc by \citet{Sicilia-Aguilar2012}. We adopt the value given by the star-count in this paper. Suppose that half of the total mass of the CrA-cloud (665~\MSun) is within this radius, the escape velocity at this distance from the center is 2.8 \kms. On the other hand, the velocity dispersion (obtained from the scatter in Gaia proper motions, assuming that the spread in RVs is the same as in RA and Dec) is $\sim$1.74~\kms. Since the velocity spread is well below the escape velocity, the system looks bound and should technically be a cluster \citep{Sicilia-Aguilar2011, Sicilia-Aguilar2013}. 

Alternatively, we may use the virial theorem to derive the mass assuming the system is bound and in equilibrium. We adopt here a radius equal to the half star-count (0.75~pc), and assume that the velocity spread represents the typical velocity of stars and cloudlets. In this way, we obtain a half mass of 440~\MSun for the core of the CrA interstellar cloud (M$_{vir}$=880~\MSun). Since only about 35~\MSun are in stars within this region, we obtain that $\sim$8\% of the interstellar matter mass has been transformed into stars at present.
This value is in good agreement with the usual efficiency of star formation in clouds. 
In fact, the star formation efficiency that is defined as $\epsilon_{\rm ff} = (SFR/M_{\rm gas}) \times t_{\rm ff}$, where $t_{\rm ff}$ is the free fall time, is found to be almost universal \citep{Krumholz2007}. 
According to \citet{Pokhrel2021}, $\epsilon_{\rm ff} =0.026$. 
Calculating $t_{\rm ff}=0.51$ Myr for the core of R CrA (we used here the virial half mass of 440~\MSun and 0.75~pc as radius), then we obtain for the star formation rate the value $SFR=22.5$~\MSun/Myr. This is in reasonable agreement with the observed mass in stars of CrA-Main if we assume that star formation lasts for 3-4~Myr. 
We expect that within a few Myr, CrA-Main will lose most of the existing gas. As a consequence, the system will become unbound, being a typical example of infant mortality among clusters. However, the crossing time for CrA-Main is $\sim~10^5$~ yr, which is much less than its age and the gas removal time, implying that we are in the adiabatic regime of gas removal \citep{Lada2003}.

\subsection{CrA-North as an expanding association}
\label{subsection:CrA-North_exanding}

\begin{figure}
   \centering
   \includegraphics[width=8.0cm]{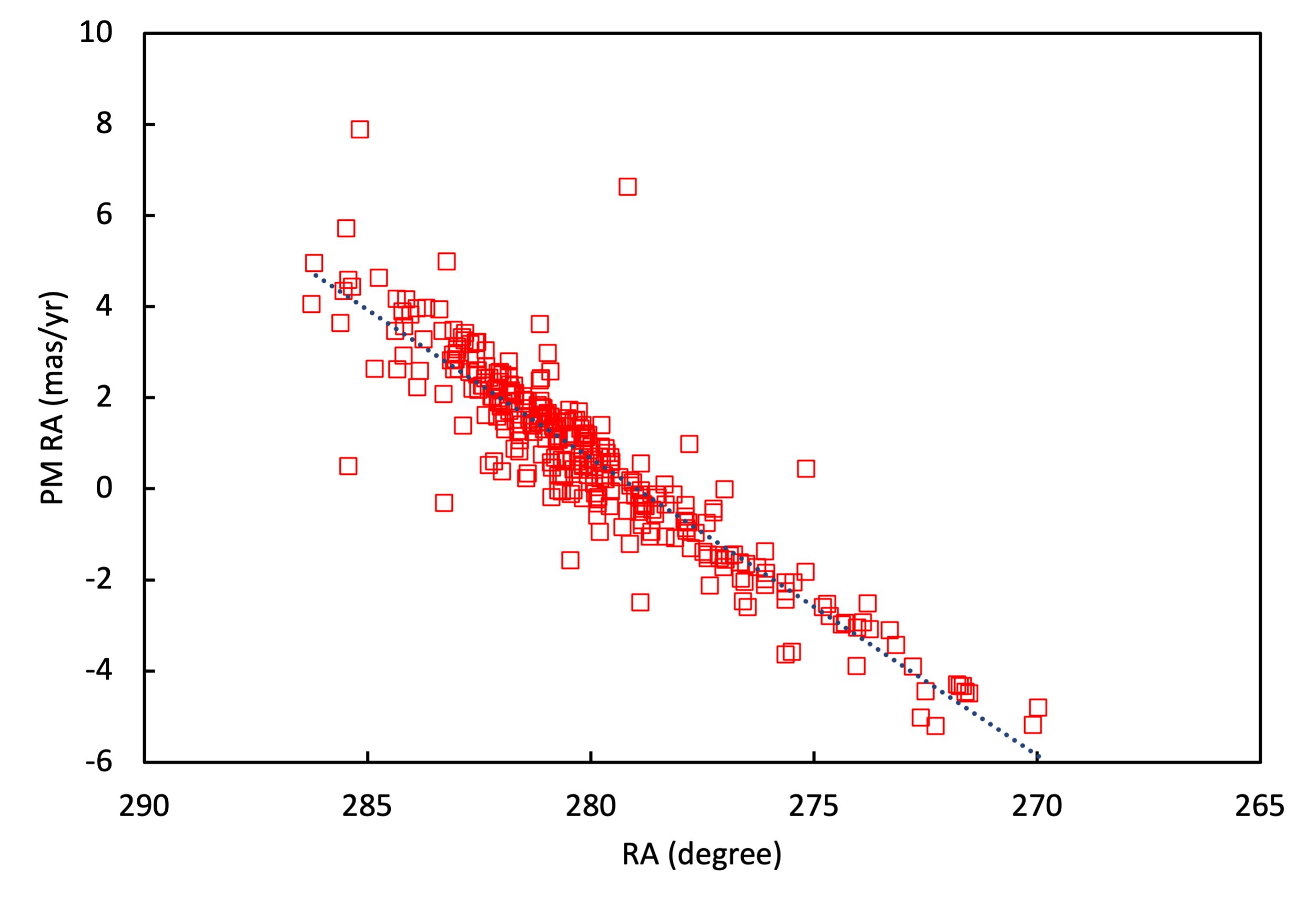}
    \caption{Proper motion along RA as a function of RA for members of CrA-North. The dashed line is the best fit line }
         \label{Fig:expansion_ra}
\end{figure}

\begin{figure}
   \centering
   \includegraphics[width=8.5cm]{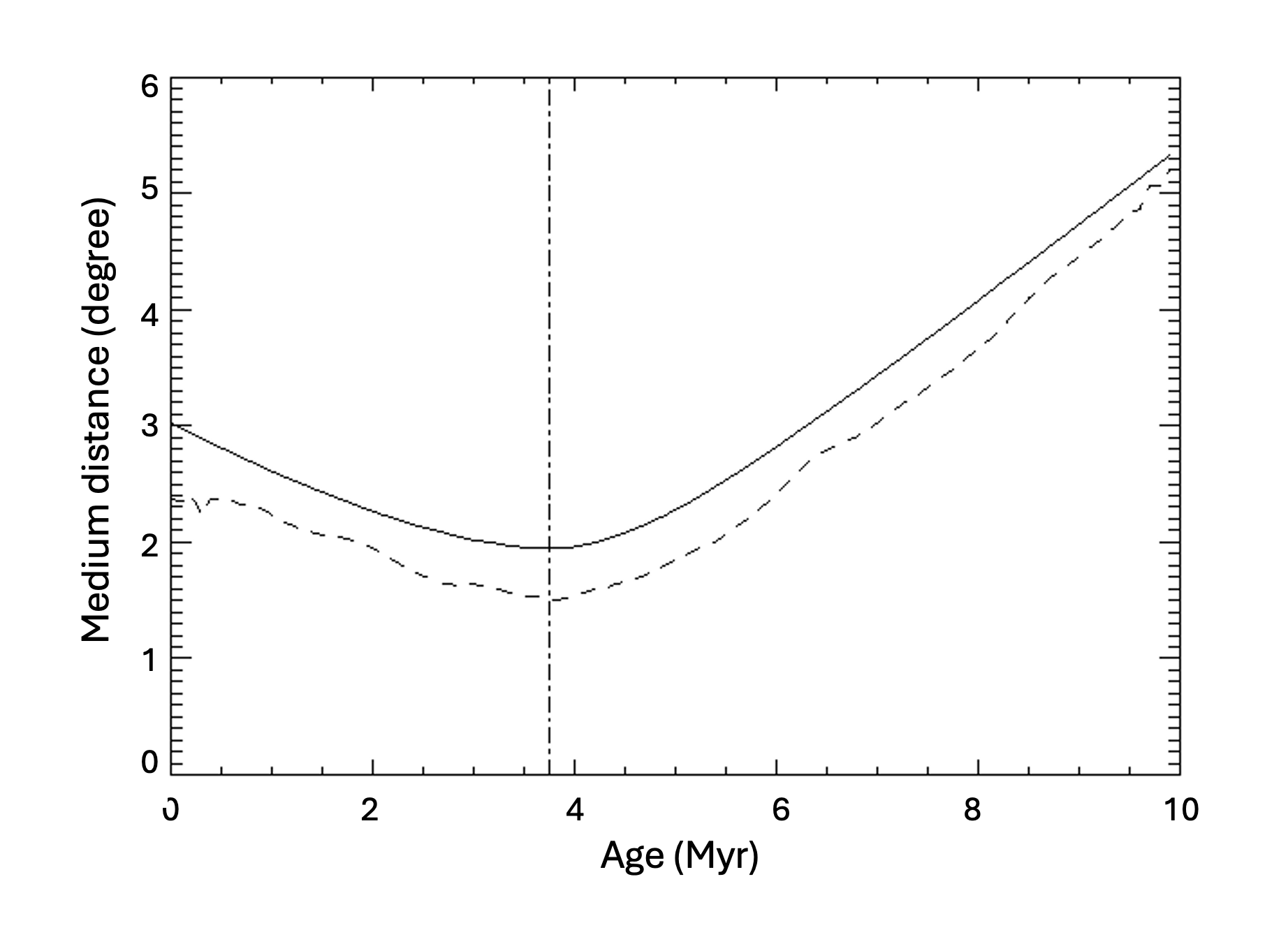}
    \caption{Run of the median apparent separation on sky of members of CrA-North as a function of time (in the past). The solid line takes into consideration the distance to all other stars in CrA-North; the dashed line only takes into consideration the distance to the closest neighbor. This last has been multiplied by 15 to have a similar vertical scale. The dash-dotted vertical line marks the position of the minium at 3.72 Myr }
         \label{Fig:expansion}
\end{figure}

There is no evidence for the presence of interstellar matter in the direction of CrA-North. The total mass in stars in CrA-North, obtained by the sum of the stars analyzed in this paper, is 130~\MSun. Lacking the contribution of interstellar matter, CrA-North is an unbound association. 
Figure \ref{Fig:expansion_ra} reports the very narrow relation between the proper motions in the right ascension and the right ascension for all stars in CrA-North, showing that CrA-North is expanding.
We notice that the right ascension roughly corresponds to the galactic latitude in the direction of the CrA complex. 
Assuming a velocity of $\sim$5~mas/yr, and a distance of a $\sim$6~degree in right ascension, we find that the cluster was more compact (with a lower scatter in RA) $\sim$4 Myr ago. Since it is possible that originally the association was not aligned N-S, a more accurate estimate of the expansion is obtained by minimizing the median distance between stars or the median distance of the closest neighbor to each star. This is obtained for an age of $3.72\pm 0.01$ Myr, as illustrated in Figure \ref{Fig:expansion}. The error bar is obtained by repeating the minimization 100 times after summing to each value of the proper motion in right ascension and declination a Gaussian distributed random value with zero mean and standard deviation equal to the median error in the proper motion of Gaia in the appropriate coordinate.

We should remind here that the expansion age of an association describes the time elapsed since gas dispersal, under the assumption of a free expansion not slowed down by self-gravity and not affected by the galactic potential. Indeed, self-gravity and the galactic potential might partially offset the time before gas dispersal. However, the expansion age is probably a lower limit to the actual age of the stars and is then consistent with the isochronal age of CrA-North we obtained in Section \ref{subsection:ages}  ($6.7\pm 0.3$ Myr).

We also notice that the CrA-North association appeared to be very elongated N-S at the epoch of its minimum size, as shown in Figure \ref{Fig:mass_pos}, which gives the position of the stars 3.72 Myr ago. The total length is $\sim 5$ degrees, that is $\sim$13~pc. Since the N-S direction roughly corresponds to galactic longitude, this apparent elongation may be attributed to a stretching due to the tides related to the galactic potential that spread stars over time along an arc along galactic longitude that grows almost linearly with time. A comparison with results obtained using publicly available code \texttt{galpy} \citep{Bovy2015}\footnote{\url{https://docs.galpy.org/en/v1.9.0/}} shows that about 14 Myr are required to cause an elongation along galactic longitude as large as that observed for CrA-North, starting from a very compact structure with a size of less than 2 pc. This agrees with the idea that Cra-North originated close to the Sco-Cen association \citep{Posch2023} at about that epoch.

\subsection{Formation history of the CrA complex}
\label{subsection:history}

\begin{figure}
   \centering
   \includegraphics[width=8.5cm]{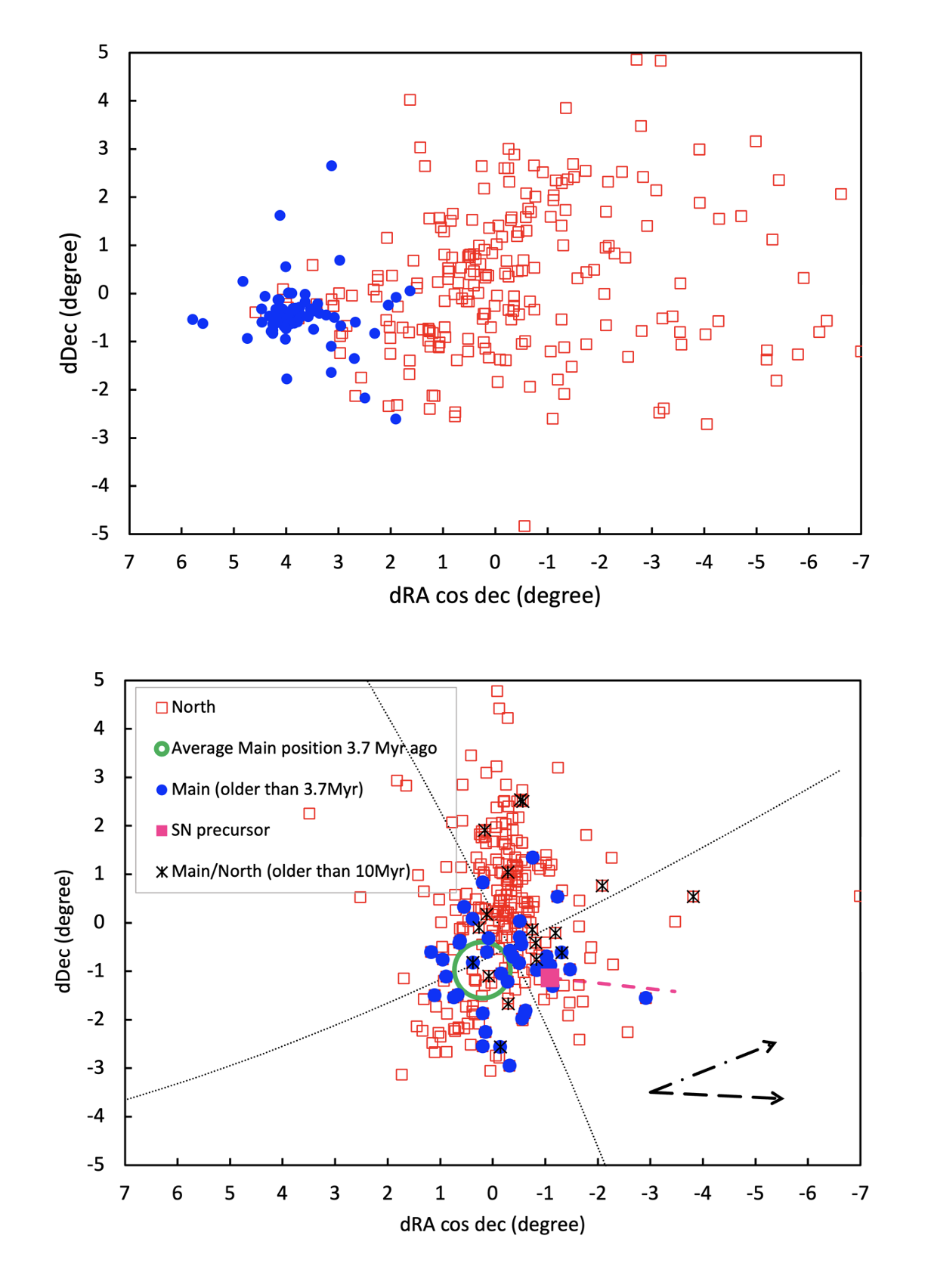}
         \caption{Position on sky of stars in the CrA-complex at present (upper panel) and 3.72 Myr ago (lower panel) relative to the center of CrA-North. 
         As labeled, red open squares are members of CrA-North, blue filled circles are members of CrA-Main (only members older than 3.72 Myr are shown in the bottom panel). In the bottom panel, the thick green circle is the average position of the CrA-Main cloud 3.72 Myr ago. The black asterisks are the positions of stars older than 10 Myr. The magenta square represents the expected location for the possible precursor of the SN that might have given the observed kick to the progenitor cloud of CrA-Main, with the dashed black line representing alternative locations. 
         The dashed lines at the bottom right indicate the direction of the Lupus and Upper Scorpius complexes. The orientation of galactic coordinates is also plotted as dotted lines for reference.}
         \label{Fig:mass_pos}
\end{figure}
In this subsection, we sketch a possible view of the history of the CrA-complex. The general framework we consider is the scenario for the formation of associations along chains of clusters outlined by \citet{Posch2023, Ratzenbock2023b, Posch2025}. In this scenario, explosions of supernovae (SNe) in the large Sco-Cen association push small gas and dust clouds ($\sim$4000~\MSun (see Appendix~\ref{Appendix_c}), approximate estimate of the sum of CrA-North and CrA-Main) far from the center of that association at a velocity of $\sim$5~km/s \citep{Posch2023}, and trigger star formation episodes in these smaller clouds. This scenario may well explain various features relative to the CrA-complex: the motion off the galactic plane, the fact that about 15 Myr ago it was close to the UCL (Upper Centaurus-Lupus) complex and 13 Myr ago to the V1062 Sco one, and the shape of the CrA-cloud that closely resembles a bow shock. However, we believe that, at least in the special case of the CrA complex, this scenario requires some modification, as discussed below.

We start noticing that CrA-Main is younger than CrA-North and that the center of Cra-Main is currently $11.5\pm 0.4$ pc from the center of CrA-North. However, combining positions and velocities to determine the variation in relative distance of the centers of Main and North with time, we found that Main and North might have been much closer in the past. We estimate the epoch of minimum separation between CrA-Main and CrA-North employing a Monte Carlo procedure. We considered Gaussian distributions for positions and proper motions for both CrA-Main and CrA-North considering the median values and standard deviations of the mean over all the members of the two groups. We then determined the epoch of minimum separation for each random extraction and repeated the procedure 30,000 times. We determined that the minimum separation of 2.64~pc (essentially in declination) occurred $3.86\pm 0.23$ Myr ago; this is compatible within the errors with the expansion age of CrA-North.
In Figure~\ref{Fig:mass_pos} we show in the top panel the position of the stars in CrA-North and CrA-Main as they appear now, and in the bottom panel the position of CrA-North and CrA-Main as they appeared 3.72 Myr ago. For CrA-Main we only plot the stars with ages older than 3.72 Myr, and the median location of the progenitor cloud of CrA-Main at that epoch is shown as a green circle. 
Following the approach of \citet{Posch2023}, we may think that the motion of CrA-Main with respect to CrA-North is the consequence of a kick received at that epoch, following which the cloud from which CrA-Main formed started moving at a velocity of 2.5~\kms along RA (toward east), 0.3~\kms (toward north) and 1.2~\kms (toward us). The total speed is then 2.8 \kms. Since the total mass of the CrA cloud core is estimated to be about 1300~\MSun (see \ref{Appendix_c}), the total momentum is about 3640~\MSun~\kms. 

The typical kinetic energy of a supernova is $10^{51}$~erg \citep{Janka2012}. 
The total momentum of a typical core collapse SN is 25000 \MSun \kms (see \citealt{Walch2015}). To explain its motion, the CrA-Main cloud should have received 15\% of the supernova momentum. The maximum fraction of the momentum received by the cloud scales with the square of the distance of the SN at the epoch of the kick. To intercept such a fraction of the total momentum, the SN should have occurred at a distance equal to about 1.27 times the radius of the cloud. If the radius of the cloud was 1.5 pc and all momentum in the direction of the cloud was intercepted, then the supernova should have occurred within about 1.9 pc from the cloud precursor of CrA-Main (green circle in Fig.~\ref{Fig:mass_pos}). 
The distance of Sco-Cen at the epoch of the kick was about 80 pc \citep{Posch2023}. This is too much to explain the momentum given to the cloud because the cloud would intercept a fraction less than $10^{-4}$ of the SN momentum. Explosions of more than a thousand SNe in Sco-Cen are needed, which is not reasonable. We conclude that the SNe responsible for the kick were not members of the Sco-Cen association. Conversely, a SN in CrA-North would be in the right distance at the epoch of the kick.

The total (current) mass in stars of CrA-North is about 130 \MSun. According to the discussion in \citet{Weidner2010} the most massive star of a cluster of this size is expected to be in the range 8-15 \MSun. It is then well possible that the most massive star in CrA-North was massive enough to explode as a SN. 
Hence, we might assume that a 15 \MSun star exploded as a SN. According to PARSEC isochrones \citep{Costa2025}, the lifetime of a 15 \MSun star is 14.2 Myr. To this age we should add the time elapsed since star formation started in CrA-North (about 3.72 Myr ago), implying that this star formed about 18 Myr ago, when CrA-North was within less than 30 pc of the UCL complex \citep{Posch2023}, during the main burst of star formation in Sco-Cen \citep{Pecaut2012}. 
There are 17 additional stars in the CrA-complex for which we obtain ages older than 10 Myr. Their position 3.72 Myr ago is marked with a special symbol in the lower panel of Figure \ref{Fig:mass_pos}; not unexpectedly, at that epoch, they were less concentrated than the younger stars of Cra-North. These stars may have formed roughly simultaneously with this massive star in the same early episode of star formation, and as their more massive brothers they remained trapped in the potential well of the gas/dust cloud that later formed the stars of CrA-North and even later of CrA-Main. 

If this interpretation is correct, the precursor of the SN that pushed away CrA-Main from CrA-North should have exploded about 2 pc south of the median position of the stars of CrA-North to push CrA-Main according to its current direction of motion. This implies that it was very close to the precursor of CrA-Main maximizing the fraction of the SN momentum that was intercepted. This may well even be $\leq 10000$ \MSun \kms, corresponding to a small explosion energy.

We also notice that if this scenario is correct and there was a star with a mass of $\sim 15$ \MSun in CrA-North, the removal of gas from the low-density regions of the association could have occurred before the SN explosion. We first notice that high velocity winds ($\sim 300$ \kms) may remove about 1\% of the mass from very young low-mass stars \citep{Hartigan1995}. This gives a momentum of about 500 \MSun \kms for a stellar association with a mass of 150 \MSun such as CrA-North. This is not enough to remove the gas. However, a star of $\sim 15$ \MSun is of spectral type B0.5 \citep{Pecaut2013} and emits Lyman continuum photons at a rate of $\sim 10^{47}$ s$^{-1}$ \citep{Sternberg2003}. The total energy emitted in the Lyman continuum during the star lifetime is $4.5\times10^{51}$ erg. Although only a small fraction ($<0.1$\%) of this energy is transferred to the interstellar medium, ionizing radiation might be sufficient to disperse a large fraction of a cloud of $\sim 10^3$ \MSun before the SN explosion. 

We finally notice that a significant fraction (about 40\%) of the stars in CrA-Main appear to be rather old, with ages comparable to those of stars in CrA-North (see Section \ref{subsection:ages}). These stars may have actually formed in the same episode that formed CrA-North (in this case making up some 14\% of stars that formed in that episode), but were trapped in the potential well of the CrA-cloud core because their relative velocity was lower than the local escape velocity and are now kinematically indistinguishable from younger stars in CrA-Main. Most of these stars are projected at the edge of the cloud, where there is only moderate interstellar absorption. They are then over-represented in Gaia photometry. This implies that the age of CrA-Main estimated from Gaia isochrones is probably overestimated.

\subsection{CrA-complex disks}
\label{subsection:disks}

\begin{figure}
   \centering
   \includegraphics[width=8.0cm]{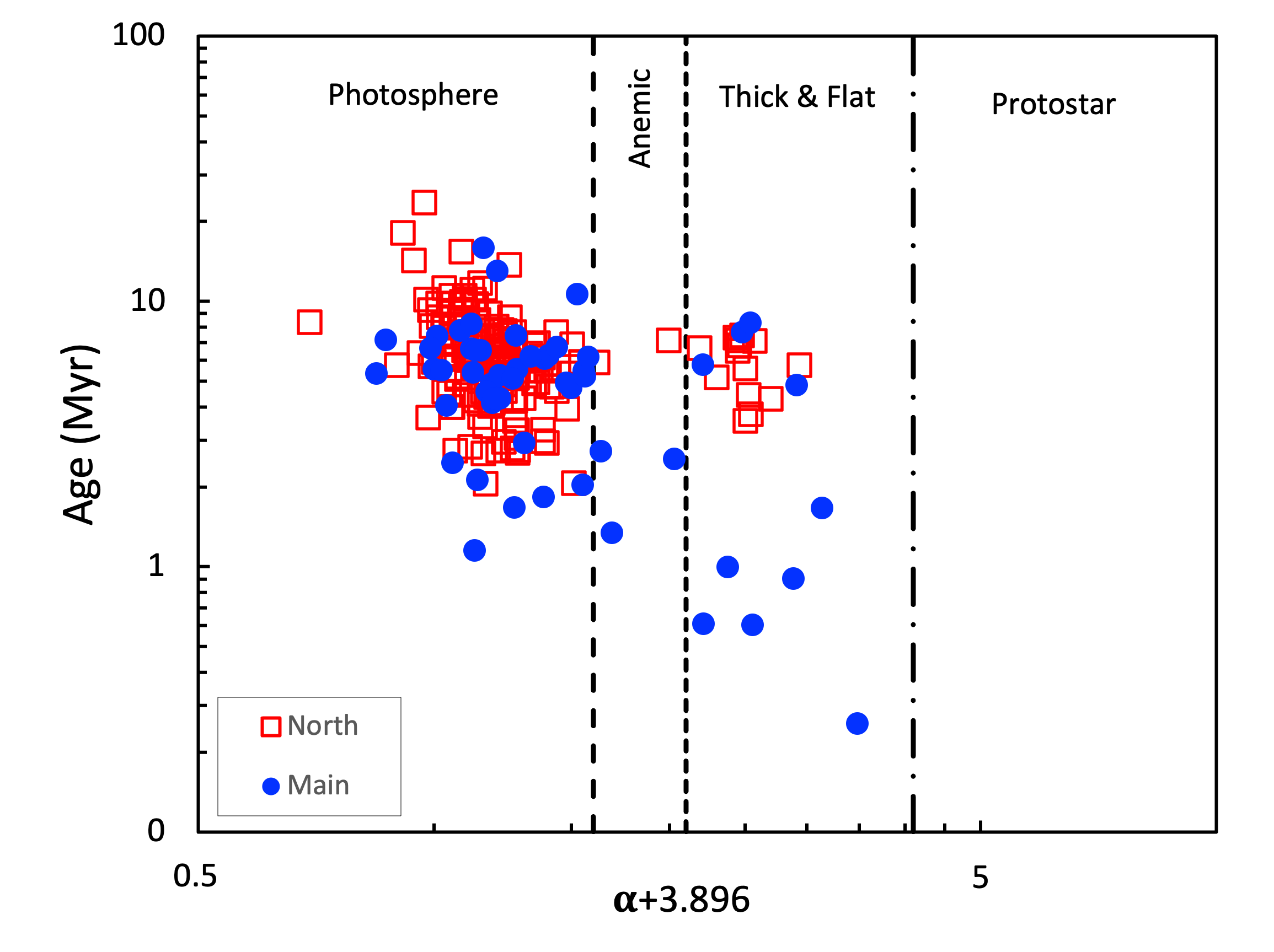}
      \caption{Age from isochrone fitting as a function of the $\alpha^{\prime}$ value (defined as the slope of the SED between W1 and W3 bands, see Sect.~\ref{section:disk}) for stars with mass $>0.075$ \MSun belonging to CrA-Main (blue circles) and CrA-North (red open squares). The dashed lines separate the different categories of disks as labeled in Table~\ref{Tab:disk_bearing_classification}. Thick disks and flat spectra are merged together.}
         \label{Fig:alpha_age}
\end{figure}

\begin{figure}
   \centering
   \includegraphics[width=8.0cm]{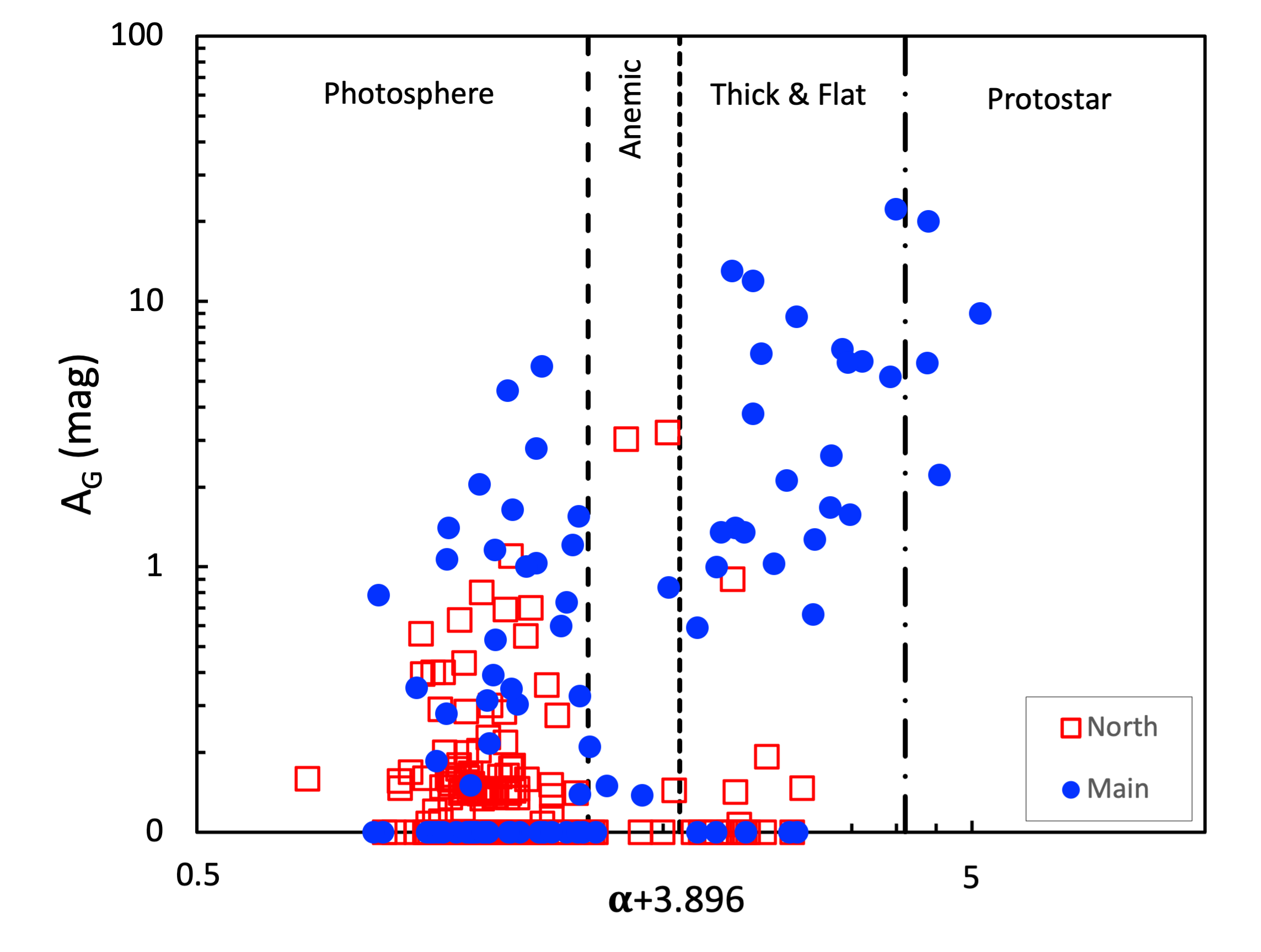}
      \caption{Interstellar absorption (A$_G$) as a function of the $\alpha^{\prime}$ value (defined as the slope of the SED between W1 and W3 bands, see Sect.~\ref{section:disk}) for stars with mass $>0.075$ \MSun belonging to CrA-Main (blue circles) and CrA-North (red open squares). The dashed lines separate the different categories of disks as labeled in Table~\ref{Tab:disk_bearing_classification}.  }
         \label{Fig:alpha_ag}
\end{figure}

\begin{table}[!ht]
\caption{Disk statistics.}
\centering
\begin{tabular}{l|cc|cc}
\hline 
Classification & \multicolumn{2}{c}{Main} & \multicolumn{2}{c}{North} \\
\hline
     & \# & \% & \# & \% \\
\hline
Photosphere   & 56 & $61\pm 5$ & 204 & $89\pm 2$ \\
Anemic        &  5 & $5\pm 2$ &   7 & $3\pm 1$ \\
Thick         & 24 & $26\pm 5$ &  18 & $8\pm 2$ \\
Flat spectrum &  3 & $3\pm 2$ &   0 &    --            \\
Protostar     &  4 & $4\pm 2$ &   0 &    --            \\
\hline
Total         & 92 &                & 229 &                \\
\hline
\end{tabular}
\label{tab:disk}
\end{table}

\begin{figure}
   \centering
   \includegraphics[width=8.0cm]{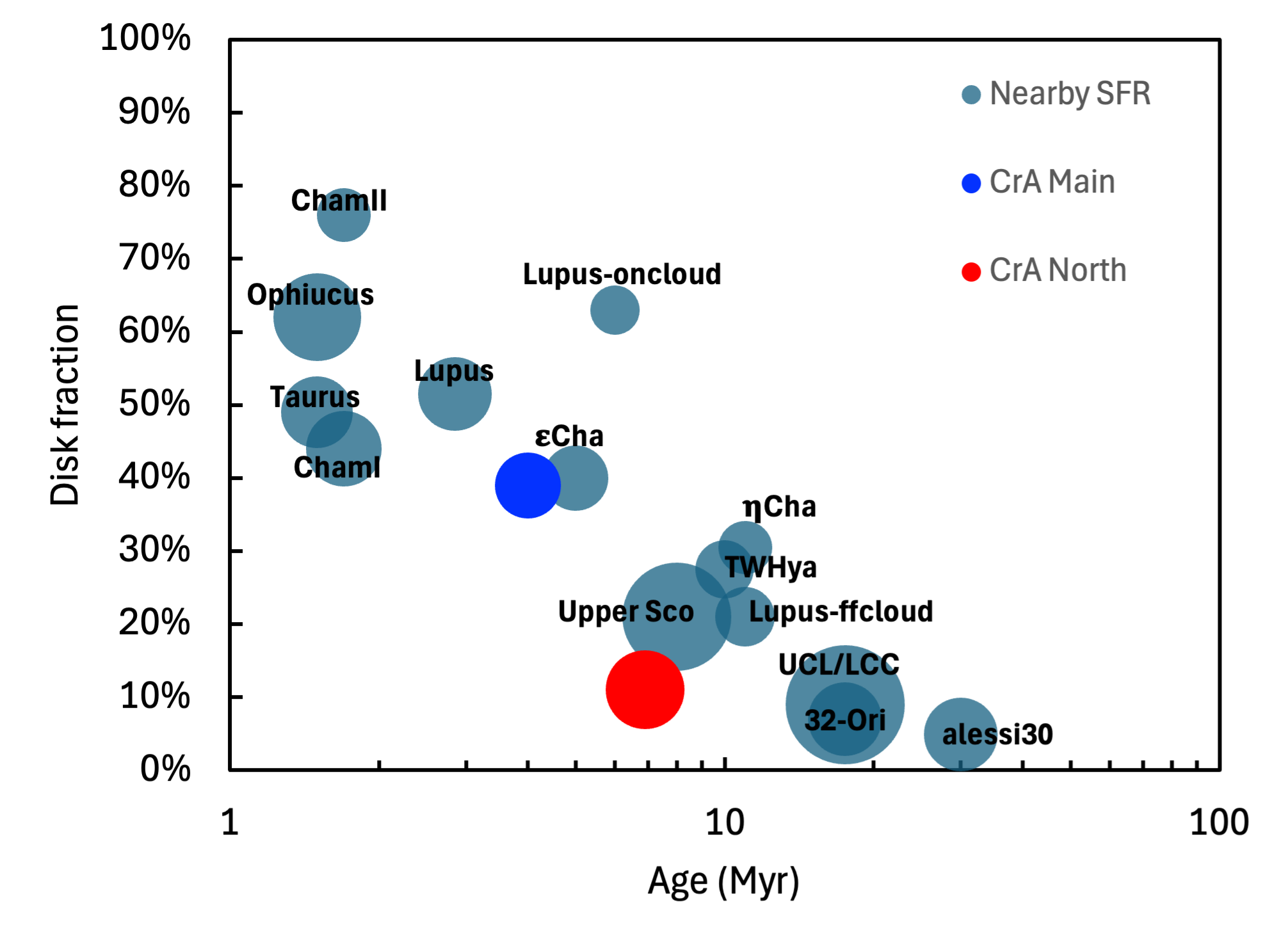}
      \caption{Frequency of disks in nearby star-forming regions and young associations (as labeled) as a function of age. Gray symbols are from \citet{Pfalzner2024}; the blue and red symbols are those we found for CrA-Main and CrA-North. The diameter of the symbols is proportional to the logarithm of the number of stars in each association. }
         \label{Fig:disk_freq}
\end{figure}

\begin{figure}
   \centering
   \includegraphics[width=8.0cm]{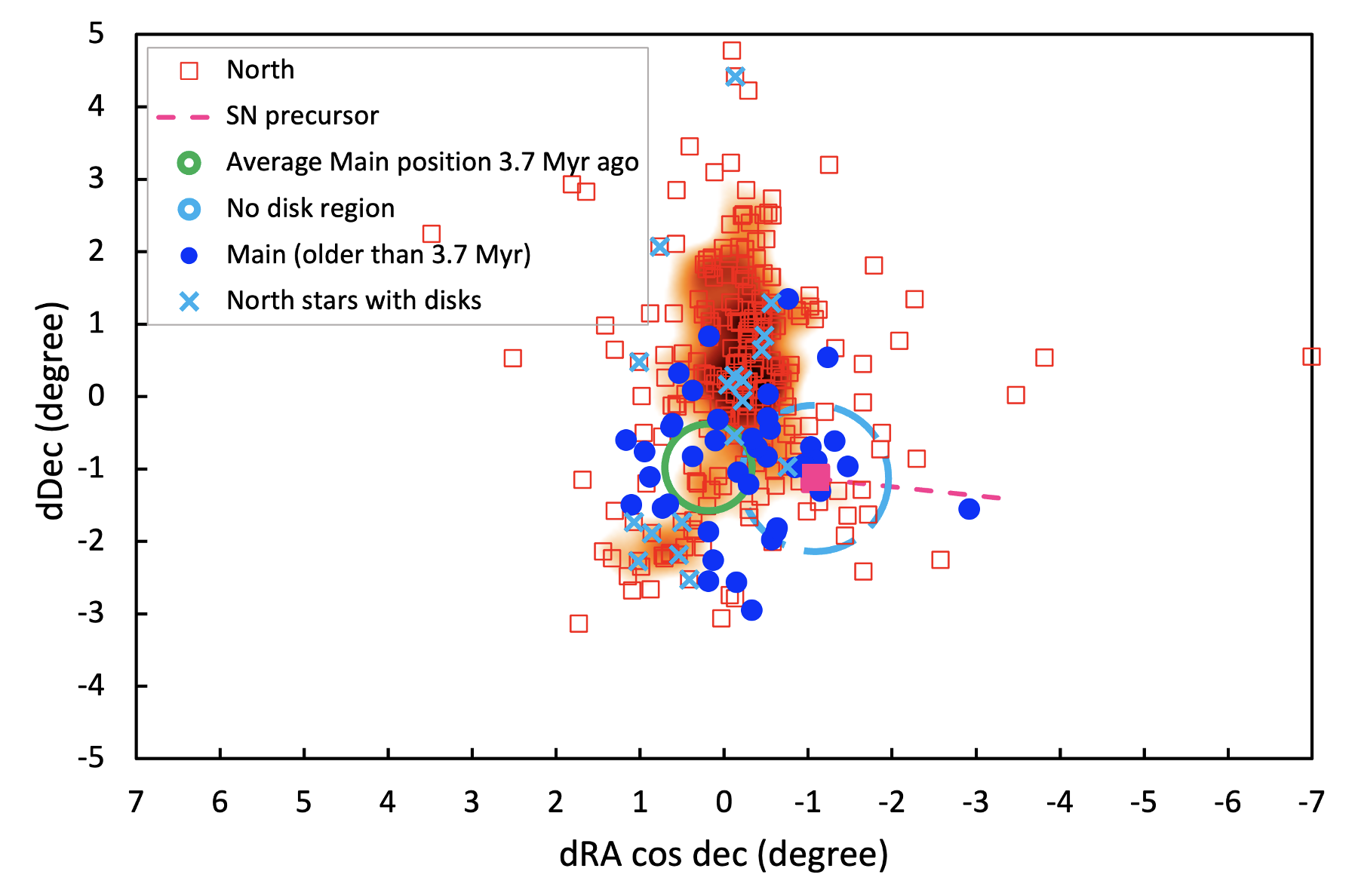}
      \caption{Position on sky of stars in the CrA-complex 3.72 Myr ago relative to the center of CrA-North. 
    Red open squares are the members of CrA-North. Cyan crosses are CrA-North stars that host long-living disks. Blue filled circles are members of CrA-Main older than 3.72 Myr. The thick green circle is the average position of the CrA-Main cloud at the epoch (3.72 Myr ago). 
    The magenta square represents the expected location for the possible precursor of the SN that might have given the observed kick to the progenitor cloud of CrA-Main, with the dashed line representing alternative locations. The light-blue dashed circle, centered on that position, has a projected radius of 1.13 degree, corresponding to 3.0 pc, which is the projected distance of the second closest disk-bearing star (cyan cross at the border of the light-blue circle). The background color is proportional to the stellar density in the CrA-complex. }
         \label{Fig:disk_pos}
\end{figure}
Table \ref{tab:disk} reports the statistics for the disks in the CrA-complex. We only consider here systems where the primary component has mass $>0.075$ \MSun. In total, disks are detected in 36 out of 92 stars in CrA-Main (that is a fraction of $0.39\pm 0.05$) and 25 out of 229 stars (a fraction of $0.11\pm 0.02$) in CrA-North. The fraction of disk-bearing stars is much higher in Cra-Main than in CrA-North, where there is no protostar or star with flat spectra. This difference is likely due to the presence of very young stars in CrA-Main. Disks are present in only four out of 38 of the stars in CrA-Main with age $>4$ Myr. This represents a fraction of $0.11\pm 0.05$ that is very similar to that observed in Cra-North, where virtually all stars are older than this limit. On the other hand, the fraction of disks is much higher in younger stars of CrA-Main: it is 10 out of 18 (a fraction of $0.45\pm 0.11$) for those with age $<4$ Myr and 22 out of 36 (a fraction of $0.61\pm 0.08$) for those missing age estimates, which mainly are located in highly reddened regions and are often very young.

To reinforce this argument, we show in Figure~\ref{Fig:alpha_age} the distribution of stars in CrA-Main and CrA-North, where on the x-axis we plot the value of $\alpha^{\prime}$ and on the y-axis the age of the stars. For stars belonging to CrA-Main, there is a clear trend of decreasing ages as we move to thicker disks. The trend is less obvious, though it is still possibly present in objects belonging to CrA-North. To show this, we only consider stars with age estimates. The mean age of the 157 stars without disks in CrA-North is $6.5\pm 0.2$ Myr, with a standard deviation of 2.9 Myr. Neglecting the few stars with anemic disks, which may be ambiguous cases, the mean age of the 15 stars with disks is $5.8\pm 0.3$ Myr, with a standard deviation of 1.3 Myr. On average, stars with disks are younger by $0.7\pm 0.4$ Myr. The test of significance for two unknown means with unknown standard deviations indicates that this result is significant at about a 95\% level of confidence. Since many stars in CrA-Main are still embedded and for those stars we do not provide an age estimate, we show a similar plot, where on the y-axis we collect the extinction in $G$ magnitude (A$_G$, see Fig.~\ref{Fig:alpha_ag}). Objects with thicker disks show higher A$_G$ values. 

We also notice that in both Figures~\ref{Fig:alpha_age} and \ref{Fig:alpha_ag}, there is a clear gap between the objects with thick disks and those with a pure photosphere. This gap is indicative of the fact that the transition between a thick disk and a pure photosphere around the star is very quick, as acknowledged by many authors (see discussion in \citealt{Alexander2014} and their Figure 1). Since WISE photometry is available for almost all stars in our sample, this implies that the estimate of the frequency of disks from these data is highly reliable.

Figure \ref{Fig:disk_freq} compares the observed frequency of stars with disks in CrA-Main and CrA-North with that found for other nearby star-forming regions and young associations (distance $<200$ pc), from \citet{Pfalzner2024}, as labeled in Figure. This quantity is plotted against age. The sizes of the symbols are proportional to the logarithm of the number of stars in each association. This is important because the frequency of disks is apparently higher in smaller star-forming regions and associations \citep{Pfalzner2024}. This figure indicates that the frequency of disks in CrA-North is about a third of that observed in other associations of similar age and size. This scarcity of disks suggests that CrA-North is an environment hostile to the survival of old disks, relative to the others considered in this plot. This result recalls the average low dust mass of disks in the CrA-complex with respect to other star-forming regions of similar age found by \citet{Cazzoletti2019}; however, the stars considered by that study are located in CrA-Main rather than CrA-North.

The frequency of disks in CrA-North is possibly slightly different in different regions of the association, being higher in the southern part and lower in the northern one. If we divide stars into three groups with the same number of members based on declination at the epoch when the expansion started (3.72 Myr ago, as discussed in Sect.~\ref{subsection:history}), the disk frequencies are $0.10\pm 0.03$ in the southern part, $0.08\pm 0.03$ in the central part, and $0.04\pm 0.03$ in the northern part. For comparison, the fractions of stars younger than the average value are $0.47\pm 0.07$, $0.59\pm 0.06$, and $0.45\pm 0.08$. The correlation is weak and cannot by itself justify the scarcity of stars in CrA North.

In principle, the scarcity of long-living disks in CrA-North can be related to a peculiarly high original density of this environment. However, there is no evidence that the original density of stars in CrA-North was higher than the current density of stars in CrA-Main, where the frequency of disks is not lower than expected for its age. To show this, we may consider the median projected distance to the closest neighbor $r$ as a measurement of the stellar concentration. This is related to the surface density $\sigma$ by the relation $r=1/(2 \sqrt{\sigma})$. We find $r=0.048$ degree for Cra-Main (that is 0.13 pc at the distance of CrA-Main; this corresponds to a median surface density of $\sigma=107$ stars/square degree) and $r=0.096$ degree for CrA-North (0.25 pc, $\sigma=27$ stars/square degree) when the association was most compact 3.72 Myr ago. In addition, the stars hosting long-living disks were preferentially located in the highest density regions of CrA-North at that epoch. This is shown in Figure \ref{Fig:disk_pos} that gives the position on the sky of stars in the CrA complex 3.72 Myr ago relative to the center of CrA-North; this age is the epoch of the kick received by CrA-Main, possibly due to the explosion of a SN (see Section \ref{subsection:history}). Different symbols are used for disk-bearing stars and those for which no disk is detected at present. 

If the scarcity of disks in CrA-North is rather due to the presence of a (now defunct) massive star in the association, it is possible that the few long living disks were either around stars that were more distant from it or were screened by interstellar dust and gas. The black symbol in Figure \ref{Fig:disk_pos} represents the expected location for the possible precursor of the SN, although we cannot exclude the fact that it was somewhat west of that position along the direction indicated by the black-dashed line. 
Figure \ref{Fig:disk_pos} shows a disk-hosting star (2MASS J18500856-3715337, cyan cross within the light-blue circle) projected at 0.37 degrees and a second (2MASS J18481012-3850421, cyan cross at the border of the light-blue circle) at 1.13 degrees from the expected position of the SN. These values correspond to 1.0 and 3.0 pc (the radius of the light-blue circle in the plot). They were both quite close to the center of the Cra-Main proto-cloud and could have been screened by interstellar matter. Ignoring this, there are 44 stars that were projected on-sky closer than 1.13 degrees to the possible SN at that epoch. Given the current frequency of disks in CrA-North, the probability that only one of these stars hosts a disk due to random fluctuations is 13\%. This is not significant, but 1.13 degrees is actually a lower limit to the distance from the possible precursor of a SN that we may consider. If we rather assumed that this was at the end of the black dashed line shown in Figure \ref{Fig:disk_pos}, 2MASS J18500856-3715337 and 2MASS J18481012-3850421 are still the closest projected disk-bearing stars, but they would have been at 2.74 and 3.45 degrees 3.72 Myr ago, corresponding to 7.2 and 9.0 pc, respectively. In this case, there would be 73 stars projected closer to the SN than 2MASS J18481012-3850421; the probability that only one of them hosts a disk due to random fluctuations would only be 1.8\%. Although not conclusive, this result suggests that stars that were close to that position at that epoch might have a lower probability than average of hosting a disk. On the other hand, this figure suggests that if there was really a massive star that exploded as a SN in CrA-North, disks only survived until the present if they were located at more than a few pc from it.

\section{Conclusions}
\label{section:conclusions}

By conducting a detailed analysis of the stellar and disk properties of the stars belonging to the CrA-Main and CrA-North groups, we have been able to provide a plausible scenario for the formation history of the nearby and isolated CrA-complex. Before highlighting the proposed scenario, we summarize here our main findings. 

\begin{itemize}
    \item CrA-Main is overall younger than CrA-North (see Fig.~\ref{Fig:histo_age}), as already noted by others \citep{Ratzenbock2023a,Ratzenbock2023b,Posch2023,Galli2020}. However, we find on average younger ages for the two groups than previously suggested (these findings are  summarized in Table~\ref{tab:summary}). Moreover, we notice that the age of CrA-Main is most likely an upper limit, since in our estimate we are missing all stars younger than $\sim$2~Myr that are still heavily embedded in the CrA-cloud. 
    \item CrA-Main is bound, while CrA-North is an unbound association that is expanding (see Fig.~\ref{Fig:expansion_ra}). A traceback analysis of the smaller distance between the stars belonging to CrA-North shows that the group was in a more compact configuration 3.72$\pm$0.01~Myr ago. As shown in Fig.~\ref{Fig:mass_pos}, CrA-North was elongated in the N-S direction at that epoch. The time it takes for the tides of the galactic potential to cause the observed elongation at 3.72~Myr starting from a compact group of $\sim$2~pc is about 14~Myr.  This value agrees with the idea proposed by \citet{Posch2023} that CrA-North originated close to the Sco-Cen association. 
    \item CrA-Main and North might have been much closer in the past than how they appear now, with a minimum separation (along declination) of 2.64~pc. 
    \item The fraction of disk-bearing stars is much higher in CrA-Main than in CrA-North (these findings are also summarized in Table~\ref{tab:summary}). Moreover, the frequency of disks in CrA-North is lower than in associations of a similar age, which suggests that CrA-North is a hostile environment for the survival of the disk. 
\end{itemize}

\begin{table}[!ht]
\caption{List of CrA-Main and CrA-North properties discussed in the paper.}
\centering
\begin{tabular}{l|c|c|c}
\hline 
Properties & CrA-Main & CrA-North & Ref. Section \\
\hline
Members & 134 & 308 & Sect.2 \\
\hline
Age   & 5.2$\pm$0.5 Myr  & 6.7$\pm$0.3 Myr & Sect.3.3 \\
\hline
Multiplicity & 89\% & 50\% & Appendix B \\
(M$>$0.8\MSun) & & & \\
\hline
Disk frequency & 39\% & 11\% & Sect.4.4 \\
\hline
Mass in stars & 69M$_\odot$ & 130M$_\odot$ & Sect.4.1/4.2 \\
\hline
\end{tabular}
\label{tab:summary}
\tablefoot{The last column contains the section where the properties are discussed. }
\end{table}
In summary, we propose that the history of the CrA complex started between 15 and 18 Myr ago, when a cloud of $\sim 4000$\MSun (approximate estimate of the mass budget in the Main+North group, see Appendix~\ref{Appendix_c}) close to the UCL region received a kick of $\sim 5$~\kms (momentum of  $2\times10^4$ \MSun \kms, \citealt{Posch2023}) from the explosion of one or more nearby SNe. This kick caused a motion of the cloud and a compression that triggered star formation. We hypothesize that the most massive star formed in this episode had a mass of about 15 \MSun. Most of the material remained in the form of a gas cloud that, with time, was stretched along galactic longitude by the galactic potential.
There was a quiescent phase with little star formation in the CrA-complex between this early SN explosion $\sim$15 Myr ago that was responsible for the separation of the CrA-complex from the main body of the UCL region\footnote{It is possible that (many) stars actually formed in this early phase but they are not any more recognized as part of the CrA-complex.}.
In a second episode of star formation about 7 Myr ago, it formed the stars that are currently part of CrA-North (a total of about 130~\MSun in stars). However, most of the mass remained in the form of a dense gas cloud of $>1,000$ \MSun. About 3.72 Myr ago, the 15 \MSun star exploded as a SN about 2 pc south of the center of CrA-North. This pushed what remained of the cloud away from CrA-North at a velocity of $\sim 2.8$ \kms and triggered a new episode of star formation, which produced the CrA-Main star-forming region. 
This is illustrated in Figure \ref{Fig:mass_pos} where we mark the possible position of this SN at an age of 3.72 Myr ago, although the SN could be west of the location indicated along the line shown in the figure. 

Intriguingly, the SN considered in this scenario matches the current position (within a projected distance of $\sim 8$~pc) and age (within 0.01 Myr) of the bright pulsar RX J1856.5-3754, which is the second closest known neutron star from the Sun \citep{Posselt2009}, in view of the fact that this explosion should have occurred very far from this position if its age is correct, considering the large value (332 mas/yr) of its proper motion \citep{Walter2002}. 
The local birth rate of pulsars is $4.7\pm 0.5$ kpc$^{-2}$ Myr$^{-1}$ \citep{Xie2024}. The probability of observing by chance a pulsar as close to the expected position, and with an age difference within 0.01 Myr from that expected for the SN required to provide the observed kick, is less than $10^{-5}$. 
A possible way out of this oddity is that RX J1856.5-3754 only acquired its high velocity in the recent past, possibly due to the disruption of a multiple system after an originally lower mass companion also exploded as a SN. 
We can provide loose limits to the time within which this disruption should have occurred. Indeed, since no extended SN remnant is observed \citep{Green2025}, this should have happened more than 30,000 years ago \citep{Reynolds2008}. However, the probability that the ejection of RX J1856.5-3754 occurred in a direction almost coincident with CrA-Main is low if the disruption event occurred more than 100000 years ago.

\section{Data Availability}

Tables~\ref{tab:main_TAB_1} and \ref{tab:main_TAB_2} are only available in electronic form at the CDS via anonymous ftp to cdsarc.u-strasbg.fr (130.79.128.5) or via http://cdsweb.u-strasbg.fr/cgi-bin/qcat?J/A+A/.

\begin{acknowledgements}

This work has made use of data from the European Space Agency (ESA) mission {\it Gaia} (\url{https://www.cosmos.esa.int/gaia}), processed by the {\it Gaia} Data Processing and Analysis Consortium (DPAC, \url{https://www.cosmos.esa.int/web/gaia/dpac/consortium}). Funding for the DPAC has been provided by national institutions, in particular the institutions participating in the {\it Gaia} Multilateral Agreement.

Part of this work is based on observations obtained with Planck (http://www.esa.int/Planck), an ESA science mission with instruments and contributions directly funded by ESA Member States, NASA, and Canada.

This publication makes use of data products from the Wide-field Infrared Survey Explorer, which is a joint project of the University of California, Los Angeles, and the Jet Propulsion Laboratory/California Institute of Technology, funded by the National Aeronautics and Space Administration.

This paper includes data collected by the TESS mission. Funding for the TESS mission is provided by the NASA's Science Mission Directorate.

ER acknowledges support from the Large Grant INAF 2022 “YSOs Outflows, Disks and Accretion (YODA): towards a global framework for the evolution of planet forming systems” and from PRIN-MUR 2022 20228JPA3A “The path to star and planet formation in the JWST era (PATH).” The work was also partially funded with an INAF "Mini-Grant" 2022. 

We thank the anonymous referee for very useful input that have improved the quality of the paper. 

\end{acknowledgements}

\bibliographystyle{aa} 
\bibliography{aa55158-25}

\begin{appendix} 

\onecolumn
\section{Mass budget of the CrA-complex}
\label{Appendix_c}
The CrA-complex is made of two components: CrA-Main and CrA-North. CrA-Main is made of stars and ISM (that is the CrA-Cloud), while there is no evidence of interstellar matter in the direction of CrA-North. 
In the following we summarize the mass budget that has been found over the years for the CrA-cloud in particular, meaning the stars in CrA-Main and the interstellar matter in this direction. 

\citet{Cambresy1999} gives a mass of 1600 \MSun (assuming a distance of 170 pc) for the CrA-cloud from star counts. Roughly half of the mass is in low-extinction regions that will not likely lead to star formation. The actual distance is 155 pc. Correcting for this distance, the total mass of the CrA-cloud would be 1330 \MSun, and half of this mass is in the CrA-cloud core, which actually corresponds to the region where stars form. This estimate of the total mass of the CrA-cloud agrees well with that of 1000 \MSun obtained by \citet{Ackermann2012} from a calibration of the $\gamma$-ray emission produced by interactions of cosmic rays and interstellar gas. 
The mass in the densest part of the cloud ($A_K > 0.2$) found by \citet{Alves2014} is 950 \MSun, and \citet{Bresnahan2018} provide a mass estimate of 1166 \MSun if a distance of 155~pc is assumed. 
In addition, in Section~\ref{subsection:CrA-Main_infant} we show that assuming the system is in equilibrium and virial theorem applies, the mass in the central region (half mass) is 440 \MSun, giving 880 \MSun for the whole CrA-Main. 
The mass of CrA-Main is then in the range 880-1330 \MSun. 
We assume a total mass of 1300 \MSun for CrA-Main.

CrA-North presently only includes stars and does not show any evidence of the presence of interstellar matter along its direction. Using our estimates, the sum of the mass is $\sim$130 \MSun.  

The sum of the current mass in stars and gas in the direction of CrA-Main and CrA-North is then about 1500 \MSun (69\MSun in CrA-Main, 130\MSun in CrA-North and 1300\MSun in the direction of CrA-Main). This is a lower limit to the original mass of the CrA-complex because part of it can have been dispersed over a wide region since its origin. Estimates of the original mass might be obtained by considering the ISM spread over wide regions of space around the CrA-complex. 
\citet{Alves2014} gives a value of 4200 \MSun for the mass of the ISM in regions of the CrA-complex with $A_K > 0.2$ and a total of possibly 7060 \MSun considering also regions very far from CrA-Main. Alternatively, we might assume that the precursor of CrA-North transformed ISM into stars with an efficiency similar to that observed for CrA-Main, that is 69/1300=0.053. If this were true, the original mass should be (69\MSun+130\MSun)/0.053$\simeq$3800~\MSun. These estimates are very uncertain; we assume that the original mass of the CrA-complex was of the order of 4000~\MSun with a factor of two uncertainty.

\section{Data for astrometric binaries} 
\label{AppendixB:data}
In Table~\ref{tab:binaries_techniques} we report the statistics of companions as explained in Section~\ref{subsection:overall} only for systems with primaries with mass $>$0.8~$M_{\odot}$. This table is comparable with those provided in \citet{Gratton2023b, Gratton2024, Gratton2025}.

In Table~\ref{tab:astrometric} we show the data obtained for the astrometric binary stars. 

\twocolumn

\begin{table*}[!ht]
\centering
\caption{Designations, proper motion and distances of the CrA-Main and CrA-North stars.}
\small
\begin{tabular}{lcccccccccc}
\hline 
\hline
Run ID & Designation  &   Alternative name & Gaia DR3 name & PM (RA) & PM (Dec) & distance \\
    &   &   &   &    (mas/yr) & (mas/yr) & (pc) \\
\hline
CrA-Main \\
\hline
1 & J18500368-3623485 & & DR3 6731838491728165248 & 4.05 & -26.89 & 155.63 \\ 
2 & J18512223-3631432 & & DR3 6731808766248601088 & 3.53 & -26.76 & 157.37\\
3 & J18512492-3903193 & & DR3 6729058823250580224 & 2.66 & -28.72 & 157.70 \\
4 & J18520787-3641319 & UCAC4 267-175041 & DR3 6731791655097086848 & 4.68 & -28.22 & 153.19 \\
...\\
134 & J19104337-3659092 & [MR81], Halpha 17 & DR3 6719057081363106048 & 1.25 & -31.68 & 156.77\\
\hline
CrA-North \\
\hline
135 & J17595255-3808323 & &      DR3 4036945490831946368 & -4.81 & -27.08 & 150.90 \\
136 & J18002133-3744285 & &      DR3 4037022383526596096 & -5.19 & -29.62 & 140.59 \\
... \\
442 & J19045816-3650279 & &  DR3 6719208225556239104 & 4.05     & -26.53 & 161.35 \\
\hline 
Pulsar & RX J1856.5-3754 & -- & & & & 167 \\
\hline                                                                  
\hline
\end{tabular}
\label{tab:main_TAB_1}
\tablefoot{
The full machine-readable list of stars analyzed in this paper is available at the CDS, while an extract of a few lines is given here. Original Gaia DR3 parameters can be queried from the Gaia Archive.\\
}
\end{table*}

\begin{table*}[!ht]
\small
\centering
\caption{Stellar and disk properties of the CrA-Main and CrA-North stars.}
\small
\begin{tabular}{lcccccccccc}
\hline 
\hline
Run ID  & SpTy  & M$_A$ & $\alpha$ & M$_{dust}$ & Age & A$_G$   & bp-rp & M$_G$ & G-H &   M$_H$ \\
 & & (M$_{sun}$) & & (M$_{Earth}$) & Myr & mag & & mag & mag \\
\hline
CrA-Main \\
\hline
1 & --      & 0.402 & 1.139 & -- & -- & 0.0 & 2.25 & 7.06 & -- & 3.77           \\
2 & M4.5    & 0.109 & 1.403 & -- & 6.3 & 0.0 & 3.07 & 9.52 & 3.77 & 5.74 \\
3 & --      & 0.106 & 1.878 & -- & -- & 0.137 & 3.19 & 9.41 & -- & 5.54\\
4 & M1      & 0.678 & 1.130 & -- & 1.2 & 0.150 & 2.04 & 6.20& -- & 3.02\\ 
...\\
134 & --    & 0.426 & 3.485 & -- & 0.3 & 1.571 & 2.91 & 7.17 & -- & 4.81\\ 
\hline
CrA-North \\
\hline
135 & --    & 0.361 & 1.091 & -- & 7.4 & 0.18 & 2.44 & 8.15 & -- & 4.77 \\
136 & --    & 0.876 & 0.945 & -- & 14.3 & 0.17 & 1.53 & 6.28 & 2.524 & 3.75 \\
... \\
442 & M5.75     & 0.067 & -- & -- & 3.0 & 0.15 & 3.91 & 11.18 & 4.33 & 6.85\\
\hline                                                                  
\hline
\end{tabular}
\label{tab:main_TAB_2}
\tablefoot{
The full machine-readable list of stars analyzed in this paper is available at the CDS, while an extract of a few lines is given here. Original Gaia DR3 parameters can be queried from the Gaia Archive.
}
\end{table*}

\begin{table}[!h]
\centering
\caption{Binary statistics for primaries with mass $>$0.8~$M_{\odot}$. }
\begin{tabular}{lcc}
\hline 
 & CrA-Main & CrA-North \\
 \hline
Total & 17 & 34 \\
\hline
\multicolumn{3}{c}{Technique}\\
\hline
Gaia & 15 (0.88) & 34 (1.0) \\
HCI & 15 (0.88) & 13 (0.38) \\
RUWE & 13 (0.76) & 34 (1.0) \\
RV & 5 (0.29) & 26 (0.76) \\
\hline
\multicolumn{3}{c}{Detection method}\\
\hline
Visual & 13 (0.76) & 14 (0.41) \\ 
RV/EB & 3 (0.18) & 0 (0.00) \\
Astrometry & 4 (0.31) & 4 (0.12) \\
\hline
\end{tabular}
\tablefoot{ In the "Technique" subtable we report the number of stars observed with the different techniques (in parenthesis the frequency with respect to the total sample). In the "Detection method" subtable we report the number of stars detected with the different methods (in parenthesis the frequency with respect to the total sample).} 
\label{tab:binaries_techniques}
\end{table}

\begin{table*}[!ht]
\centering
\caption{Data for astrometric binaries.}
\begin{tabular}{lcccccccccc}
\hline
\small
2MASS & RUWE&RV&error&$N_{Obs}$&Ampl.&S/N(PMa)&$M_A$&$M_B$&$a$&$q$\\
&&\kms&\kms&&\kms&&\MSun&\MSun&au&\\
\hline
\multicolumn{11}{c}{CrA-Main}\\
\hline
J18500368-3623485& 8.448&  -2.81& 7.24&11& 63.82&    &0.402&0.331& 0.79&0.823\\
J18512492-3903193& 9.349&  -1.87&     &  &      &    &0.106&0.082& 1.65&0.774\\
J18580182-3653451& 1.127& -15.99& 6.35&12& 60.20&    &0.638&0.225& 0.04&0.353\\
J18593179-3650261& 3.567&  -2.56& 1.81&13& 15.37&    &0.347&0.205& 1.88&0.591\\
J19003158-3652136& 1.467&  -2.54&     &  &      &    &0.099&0.036&29.77&0.364\\
J19004728-3647467& 1.668&  -2.56&     &  &      &    &0.140&0.071& 3.32&0.507\\
J19010325-3703392& 0.871&   0.90&     &  &      &4.19&2.750&0.234& 8.00&0.085\\
J19010971-3647528& 1.240&  -4.15& 3.14&12& 33.08&    &1.118&0.240& 0.04&0.215\\
J19012872-3659317& 2.118&  -6.39& 8.48& 9&      &    &0.334&0.184& 3.34&0.551\\
J19012901-3701484& 2.106&  -6.39& 2.08& 9& 13.74&    &0.163&0.066& 1.70&0.405\\
J19013357-3700304& 1.516&  -2.48&     &  &      &    &1.412&0.577& 7.73&0.409\\
J19015523-3723407& 6.134&  -2.34&     &  &      &    &0.734&0.547& 2.24&0.745\\
J19023308-3658212& 7.913&  -2.29&     &  &      &    &0.304&0.228& 2.45&0.750\\
J19031609-3714080& 2.183&  -6.83&     &  &      &    &0.210&0.120& 2.96&0.571\\
J19094592-3704261& 2.102&  -2.35&17.55&11&160.81&    &0.978&0.698& 0.09&0.714\\
\hline  
\multicolumn{11}{c}{CrA-North}\\                
\hline
J17595255-3808323& 1.681&  -7.39& 4.53&13&      &    &0.361&0.203& 2.74&0.562\\
J18065523-3753495& 4.634&       &     &  &      &    &0.367&0.205& 1.86&0.559\\
J18102445-3700556& 1.690&  -6.08&36.95&18& 19.61&    &0.750&0.338& 3.11&0.451\\
J18160486-3738004& 5.433&-134.94& 5.60& 6&      &    &0.163&0.145& 1.72&0.889\\
J18183306-3450488& 1.485&       &     &  &      &    &0.350&0.135&17.98&0.385\\
J18275948-3713550& 6.449&       &     &  &      &    &0.661&0.511& 2.16&0.773\\
J18280442-3258331& 1.787& -11.24& 4.80& 9& 36.08&    &0.565&0.332& 1.03&0.588\\
J18310898-3527474& 5.108&   2.40& 5.24& 5&      &    &0.233&0.177& 1.94&0.758\\
J18312455-3706401& 7.072&       &     &  &      &    &0.149&0.132& 1.80&0.885\\
J18312648-3454402& 1.882&       &     &  &      &    &0.056&0.034& 2.07&0.603\\
J18331574-3730147& 4.729&       &     &  &      &    &0.479&0.389& 2.15&0.812\\
J18343120-3345559& 4.442&  -2.22& 5.03&12&      &    &0.157&0.126& 1.89&0.803\\
J18351241-3235539&17.877&       &     &  &      &    &0.462&0.407& 2.02&0.882\\
J18363954-3451257& 1.788&  -6.04& 7.76&13& 77.98&    &0.539&0.518& 0.11&0.961\\
J18385839-3508514& 4.611&       &     &  &      &    &0.217&0.171& 1.95&0.787\\
J18392259-3558365& 2.841&       &     &  &      &    &0.203&0.185& 1.99&0.911\\
J18404036-3606088& 1.660&  -0.67& 2.99&17&      &    &0.471&0.255& 3.01&0.542\\
J18405612-3750171& 1.547&       &     &  &      &    &0.405&0.189& 3.76&0.467\\
J18413826-3502327& 1.412&       &     &  &      &    &0.177&0.043&44.82&0.245\\
J18414669-3629339& 1.620&  -3.07& 1.61&11& 13.29&    &0.849&0.368& 0.62&0.433\\
J18415143-3807380& 1.433&       &     &  &      &    &0.051&0.015&42.88&0.293\\
J18425245-3735568&12.772&       &     &  &      &    &0.430&0.348& 2.21&0.810\\
J18430753-3658293&15.217&       &     &  &      &    &0.471&0.385& 2.13&0.817\\
J18433285-3626303& 1.488&   0.11& 5.67&12&      &    &0.325&0.131&17.91&0.402\\
J18435838-3559096& 1.488&  -7.01& 3.78& 9&      &    &0.274&0.099&17.78&0.361\\
J18443116-3723347& 6.439&   1.87&11.97& 9& 92.04&    &0.830&0.618& 0.73&0.744\\
J18454492-3855055& 1.911&       &     &  &      &    &0.111&0.066& 2.03&0.591\\
J18480666-3733074& 2.178& -17.64& 6.18& 7&      &    &0.374&0.227& 2.61&0.606\\
J18481314-3712485& 1.445&       &     &  &      &    &0.257&0.074&34.52&0.286\\
J18490142-3559143& 2.100&       &     &  &      &    &0.072&0.046& 2.14&0.642\\
J18514581-3604499& 7.244&       &     &  &      &    &0.153&0.128& 1.89&0.837\\
J18520378-3847322& 1.727&       &     &  &      &    &0.492&0.260& 3.07&0.528\\
J18521730-3700119& 1.577&       &     &  &      &    &0.983&0.440& 3.11&0.448\\
J18553379-3629349& 1.833&       &     &  &      &    &0.184&0.105& 2.71&0.569\\
J18564094-3741094& 1.683&  -1.83& 5.47& 8&      &    &0.215&0.113& 2.51&0.525\\
J18591874-3551423& 1.680& -28.05& 3.72& 9&      &    &0.407&0.235& 2.73&0.578\\
J19044441-3650410& 4.445&  -1.77& 0.45&17&  5.21&4.95&2.180&0.634& 2.17&0.291\\
\hline
\end{tabular}
\label{tab:astrometric}
\end{table*}

\end{appendix}
\end{document}